%% file: main.tex
\documentclass{article}

\pdfoutput=1
\input{headerarxiv}

\begin{document}
\label{firstpage}

\maketitle

\input{abstractarxiv}

%%%%%%%%%%%%%%%%%%%%%%%%%%%%%%%%%%%%%%%%%%%%%%%%%%%%%%%%%%%%%%%%%%%%%%%%

\section{Introduction}
\label{introduction.sec}
\input{introduction}

%%%%%%%%%%%%%%%%%%%%%%%%%%%%%%%%%%%%%%%%%%%%%%%%%%%%%%%%%%%%%%%%%%%%%%%%

\section{Wave Model, Imaging Conditions and RTM Procedure}
\label{inverseProblem.sec}
\input{problem}

%%%%%%%%%%%%%%%%%%%%%%%%%%%%%%%%%%%%%%%%%%%%%%%%%%%%%%%%%%%%%%%%%%%%%%%%

\section{Numerical Schemes}
\label{schemes.sec}

In this section, we detail the numerical schemes used to solve for the source and receiver wavefields and to evaluate the imaging condition $I(\vec{x})$ from their correlation.
The spatial discretization is based on a nodal discontinuous Galerkin scheme. The time-stepping is performed through a multi-rate scheme with local time stepping where needed.
These are detailed in sections~\ref{sec:schemeDG} and \ref{sec:schemeTime}, respectively.
The numerical evaluation of $I(\vec{x})$ is addressed in section~\ref{sec:schemeRTM}.

\input{schemedg}
\input{schemetime}
\input{schemertm}

%%%%%%%%%%%%%%%%%%%%%%%%%%%%%%%%%%%%%%%%%%%%%%%%%%%%%%%%%%%%%%%%%%%%%%%%

\section{Parallel Computation and Implementation}
\label{computation.sec}

In this section, we describe the RiDG implementation for massively parallel computations on accelerator-aided clusters.
The fundamentals and features of OCCA are described in section~\ref{computationOCCA}.
Our implementation for a single multi-treading device is detailed in section \ref{computationKernels} and the adaptations for distributed-memory clusters are explained in section \ref{computationMPI}.
The memory transfers required for MPI communications and for the RTM procedure are discussed in section \ref{computationRTM}.

\input{computationocca}
\input{computationkernels}
\input{computationmpi}
\input{computationrtm}

%%%%%%%%%%%%%%%%%%%%%%%%%%%%%%%%%%%%%%%%%%%%%%%%%%%%%%%%%%%%%%%%%%%%%%%%

\section{Computational Results}
\label{benchmarks.sec}
\input{benchvalid}

\input{benchmpi}

%%%%%%%%%%%%%%%%%%%%%%%%%%%%%%%%%%%%%%%%%%%%%%%%%%%%%%%%%%%%%%%%%%%%%%%%

\section{Conclusion}
\label{conclusion.sec}
\input{conclusion}

%%%%%%%%%%%%%%%%%%%%%%%%%%%%%%%%%%%%%%%%%%%%%%%%%%%%%%%%%%%%%%%%%%%%%%%%

\section*{Acknowledgements}
AM is Honorary Fellow of the Belgian American Educational Foundation (BAEF).
He acknowledges Wallonie-Bruxelles International (WBI) for an excellence grant.
TW acknowledges generous support from the Shell Oil Company and Shell Global Solutions International B.V..
The authors acknowledge the Texas Advanced Computing Center (TACC) at The University of Texas at Austin for providing HPC resources that have contributed to the research results reported within this paper.
AM thanks Rajesh Gandham and David Medina for helpful discussions. We are grateful to Ty Mckercher and NVIDIA Corporation for providing GPU equipment.
Approved for release under RTI 95514115.

%%%%%%%%%%%%%%%%%%%%%%%%%%%%%%%%%%%%%%%%%%%%%%%%%%%%%%%%%%%%%%%%%%%%%%%%

\bibliographystyle{abbrvnat}
\bibliography{myrefs}

%%%%%%%%%%%%%%%%%%%%%%%%%%%%%%%%%%%%%%%%%%%%%%%%%%%%%%%%%%%%%%%%%%%%%%%%

\appendix
\section{Derivation of the Nodal DG Scheme}
\label{appendixScheme.sec}
\input{schemedgappendix}

%%%%%%%%%%%%%%%%%%%%%%%%%%%%%%%%%%%%%%%%%%%%%%%%%%%%%%%%%%%%%%%%%%%%%%%%%%%%%%%%%%%%

\label{lastpage}
\end{document}

%% file: headerarxiv.tex
\usepackage{fullpage}

\usepackage{amsmath,amssymb,amsfonts}
\usepackage{stmaryrd}
\usepackage{xfrac}
\usepackage{eqnarray}

\usepackage{authblk}
\usepackage{natbib}

\usepackage[ruled,vlined]{algorithm2e}

\usepackage{color}
\definecolor{gray}{gray}{0.5}
\definecolor{lightgray}{gray}{0.75}
\definecolor{lightlightgray}{gray}{0.9}

\newcommand{\diff}[2]{\frac{\partial #1}{\partial #2}}
\newcommand{\ddiff}[2]{\frac{d #1}{d #2}}

\newcommand{\dt}{\Delta t}
\newcommand{\cm}{\textrm{cm}}
\newcommand{\Hz}{\textrm{Hz}}
\newcommand{\m}{\textrm{m}}
\newcommand{\km}{\textrm{km}}
\newcommand{\g}{\textrm{g}}
\newcommand{\s}{\textrm{s}}
\newcommand{\D}{\mathsf{D}}
\newcommand{\rec}{\text{R}}
\newcommand{\sou}{\text{S}}
\newcommand{\fields}{\text{fields}}
\newcommand{\traces}{\text{traces}}
\newcommand{\pad}{\text{pad}}
\renewcommand{\dim}{\text{dim}}
\newcommand{\blkS}{\text{blkS}}
\newcommand{\blkV}{\text{blkV}}

\newcommand{\disc}[1]{{#1}}

\newcommand{\eg}{\textit{e.g.} }
\newcommand{\ie}{\textit{i.e.} }

\newcommand{\jump}[1]{\left\llbracket{#1}\right\rrbracket}

\newcommand{\mat}[1]{ {\bf #1 } }
\renewcommand{\vec}[1]{ {\bf #1 } }

\title{Nodal Discontinuous Galerkin Simulations\\
for Reverse-Time Migration on GPU Clusters}

\author[1]{A. Modave\footnote{Corresponding author: modave@rice.edu}}
\author[2]{A. St-Cyr}
\author[2,3]{W. A. Mulder}
\author[1]{T. Warburton}
\affil[1]{Rice University, Houston, Texas, USA}
\affil[2]{Shell Global Solutions International B.V., Rijswijk, The Netherlands}
\affil[3]{Delft University of Technology, The Netherlands}

\date{}

%% file: abstractarxiv.tex
\begin{abstract}
Improving both accuracy and computational performance of numerical tools is a major challenge for seismic imaging and generally requires specialized implementations to make full use of modern parallel architectures.
We present a computational strategy for reverse-time migration (RTM) with accelerator-aided clusters.
A new imaging condition computed from the pressure and velocity fields is introduced.
The model solver is based on a high-order discontinuous Galerkin time-domain (DGTD) method for the pressure-velocity system with unstructured meshes and multi-rate local time-stepping.
We adopted the MPI+X approach for distributed programming where X is a threaded programming model.
In this work we chose OCCA, a unified framework that makes use of major multi-threading languages (\eg CUDA and OpenCL) and offers the flexibility to run on several hardware architectures.
DGTD schemes are suitable for efficient computations with accelerators thanks to localized element-to-element coupling and the dense algebraic operations required for each element.
Moreover, compared to high-order finite-difference schemes, the thin halo inherent to DGTD method reduces the amount of data to be exchanged between MPI processes and storage requirements for RTM procedures.
The amount of data to be recorded during simulation is reduced by storing only boundary values in memory rather than on disk and recreating the forward wavefields.
Computational results are presented that indicate that these methods are strong scalable up to at least 32 GPUs for a three-dimensional RTM case.
\end{abstract}

%% file: introduction.tex
Reverse-time migration (RTM), introduced in the early 80s \citep{Baysal1983,Loewenthal1983,Mcmechan1983,Whitmore1983,ref:lailly,ref:tarantolaacou}, is a migration method based on the wave equation. Although not the only means of obtaining images of the subsurface from seismic data, it is currently widely used in the oil \& gas industry \citep{Etgen2010}.
RTM constructs a subsurface image by correlating source wavefields with time-reversed receiver wavefields
\cite[\eg][]{Claerbout1985,Bleistein2001}. However, the procedure is very demanding in terms of computational and memory resources. It can be implemented as a single-step procedure or as an iterative scheme, minimizing the
difference between observed primary reflections and synthetic data modelled by the Born approximation
of the wave equation.
At each iteration, the solutions of two numerical simulations have to be computed and cross-correlated:
a forward simulation to determine the source wavefields in a background model that is reflection-free
in the seismic bandwidth and a reverse-time wavefield generated by the data or data residuals at the receivers.
In typical industrial applications, both simulations are expensive and must be performed on clusters with many-core CPUs and, optionally, with accelerators such as GPUs to further boost the performance. It is therefore critical to develop numerical methods that are both efficient and scalable on such clusters. Also, strategies have to be found to compute the RTM cross-correlation without penalizing the final run time.

Nowadays, finite-difference methods are the most widely used numerical schemes. 
A large literature is available, notably to reach high-order convergence rates and to reduce dispersion and dissipation errors \cite[see \eg][for a review]{Virieux2011}.
Alternatively, several kinds of finite-element methods have been proposed, such as spectral finite-element methods \citep{Komatitsch1998, Komatitsch1999}, continuous mass-lumped finite-element methods \citep{Chin1999, Cohen2001} or discontinuous Galerkin (DG) methods \citep{Collis2010, Dumbser2006, Etienne2010, Krebs2014, Mercerat2015}.
In contrast to the standard finite-element approach, those methods do not require the solution of large sparse linear systems of equations, but if a lumping procedure or a block diagonal mass matrix is chosen,
it is possible to use an explicit time-stepping scheme.
Such methods have some advantages over finite-differences.
They can easily handle geometric or physical discontinuities by using unstructured meshes without suffering from the loss of accuracy that afflicts most finite-difference schemes.
Moreover, a hybrid discretization can be deployed thanks to the flexibility in mixing several kinds of elements and different discretization orders.
In comparative work \citep{Baldassari2012, Minisini2012, Moczo2011, Zhebel2014}, these finite-element methods generally exhibit similar performance results and they surpass classic finite-difference schemes in cases with geometric or physical discontinuities.
However, in three dimensions, spectral elements are currently only available for hexahedral meshes, which are less flexible than tetrahedral meshes.
Continuous mass-lumped finite elements are available for tetrahedrons only up to third-degree polynomial basis functions, enabling at most a fourth-order spatial convergence of the scheme.
The DG approach does not suffer from these limitations, but it is generally considered as expensive because of 
the excessively large number of degrees of freedom.
In contrast with other methods, DG schemes are easily combined with local time-stepping strategies in order to improve computational efficiency and speed up of geophysical wave propagation \citep{Baldassari2011,Dumbser2009,Minisini2013}.

Accelerator-aided clusters require specialized algorithms and computational implementations to make full use of the hardware.
In the computational geophysics community, several implementations have been proposed for the numerical modeling of seismic wave propagation problems with clusters of GPUs.
They are based on finite-difference schemes \citep{Michea2010,Weiss2013}, a spectral element scheme \citep{Komatitsch2010} or a discontinuous Galerkin method \citep{Mu2013b}.
To our knowledge, only a few papers have been published for complete RTM procedures on GPUs, and only with finite-difference schemes \citep{Abdelkhalek2012,Liu2013,Yang2014}. On the other hand, \cite{Klockner2009} have proposed GPU implementations for wave-like problems written with time-domain first-order wave systems and discretized with nodal DG schemes, but does not address RTM imaging.
The local element-to-element coupling and the dense algebraic operations required per element make nodal DG methods suitable for parallel multi-threading computations, especially with GPUs.
This implementation has successfully been adapted for several applications \citep{Fuhry2014,Gandham2015,Godel2010,Modave2015}.

A supplementary difficulty for RTM procedures is to compute the cross-correlation of the solutions of source and receivers simulations.
Because those simulations are performed in opposite time directions, it is common to either
store the source wavefield at check-points or to reproduce it backward in time, in order to compute the cross-correlation during the reverse-time simulation for the receiver wavefield.
The source wavefield can be reproduced by using a check-pointing strategy or by saving boundary values during a first forward run \citep{Dussaud2008,Symes2007}.
However, in both cases, a substantial storage space is required as well as memory transfers during the runs, which can slow down the total run time.
A strategy based on random boundaries can overcome this bottleneck, but the final image can be polluted by noise if the coverage of the survey is not dense enough \citep{Clapp2009,Liu2012}.
Several nonreconstructive imaging methods have been recently proposed by using alternative imaging conditions \citep{Nguyen2015}.

In this paper, we propose a novel implementation, named \textit{``RiDG''}, for RTM procedures with clusters of GPUs.
It is based on first-order wave systems, with a high-order nodal DG scheme \citep{Hesthaven2002,Hesthaven2007}, unstructured meshes made of tetrahedras, and a multi-rate Adams-Bashforth time-stepping \citep{Godel2010}.
Using first-order systems instead of second-order allow to investigate alternative imaging conditions.
We describe here an implementation for the pressure-velocity system and a novel imaging condition based on the knowledge of both pressure and velocity, which improves final RTM images in comparison with the classical condition.

RiDG is coded with the C++ language, OCCA for multi-threaded programming, and MPI for parallel computing with distributed memory.
OCCA \citep{Medina2014} is a unified programming framework that abstracts major multi-threading languages (OpenCL, CUDA, pThreads and OpenMP), offering flexibility to choose the hardware architecture and thread-model at run-time.
RiDG directly follows the strategies of \cite{Klockner2009} for GPU computing.
We have particularly taken care of memory transfers and memory storage that are required for a complete RTM procedure.
Despite those transfers, RiDG reaches an excellent strong scalability for a three-dimensional case with 32 GPUs.

This paper is organized as follows.
In section 2, the mathematical wave model and the novel imaging condition are introduced.
Section 3 is dedicated to numerical schemes.
The high-order nodal DG scheme, the multi-rate Adams-Bashforth time-stepping and the numerical approximation of the imaging condition are detailed.
In section 4, the computational implementation with OCCA and MPI is described.
In particular, the OCCA programming framework is introduced, and the management of computational work and memory transfer is discussed.
Finally, computational results for three-dimensional benchmarks are provided in section 5.

%% file: problem.tex
The RTM implementation described in this paper is based on the acoustic wave model formulated in terms of pressure $p(\vec{x},t)$ and particle velocity $\vec{v}(\vec{x},t)$,
\begin{eqnarray}
  \diff{p}{t} + \rho c^2 \: \nabla\cdot\vec{v} &=& f(t)\:\delta(\vec{x}-\vec{x}_0),
    \label{eqn:waveSyst1} \\
  \rho \diff{\vec{v}}{t} + \nabla p &=& 0.
    \label{eqn:waveSyst2}
\end{eqnarray}
In this work, both the density $\rho$ and phase velocity $c$ are assumed to be positive and piecewise constant.
In the source term, $f(t)$ is understood as a signal and $\delta(\vec{x}-\vec{x}_0)$ is the Dirac delta function.
%At interfaces where physical parameters are discontinuous, we assume the continuity of both pressure and the normal component of particle velocity.
In the complete mathematical model, a free-surface boundary is usually represented with the boundary condition $p = 0$,
and a transparent boundary can be approximated with the boundary condition
\begin{eqnarray}
  p - \rho c \: \vec{n}\cdot\vec{v} &=& 0.
    \label{eqn:ABC}
\end{eqnarray}

For each set of source and receiver signals, the RTM procedure requires the numerical solutions of two wave propagation problems using an initial model of the subsurface material parameters.
The `source' solution is obtained by emitting the source signal forward-in-time, and the `receivers' solution is computed by re-injecting backward-in-time the signals recorded at receivers.
The key step of the method then consists in computing an imaging condition that correlates both solutions.
A classical imaging condition \citep{Bleistein2001,Claerbout1985} is
\begin{eqnarray}
  I(\vec{x}) &=& \sum_s \int_{0}^{t_f}
    p_\sou(\vec{x},t;s)\;p_\rec(\vec{x},t;s)\;dt,
  \label{eqn:imageP}
\end{eqnarray}
where $p_\sou(\vec{x},t;s)$ and $p_\rec(\vec{x},t;s)$ are the pressure fields of the source and receivers wavefields for shot $s$, respectively, and $t_f$ is the final time. In the following will neglect the shot index $s$.
Note that equation~\eqref{eqn:imageP} is expressed in its simplest form, without ``true-amplitude'' migration weights,
as we want to focus on the computational rather than on the geophysical issues.
High values of $I(\vec{x})$ correspond to geophysical features of interest, generally called reflectors.
%A typical seismic survey involves thousands of source-receiver samples called shot gathers.
%Away from reflectors, the waves from the various shot gathers will sum up to white noise whereas, near them, the source and receiver waves will be strongly correlated and will stack together to expose the hidden reflector.

While most RTM implementations are based on the second-order wave equation that provides only the pressure field, both pressure $p$ and velocity $\vec{v}$ are available with our first-order formulation.
This allows us to investigate different definitions for the imaging condition, using both types of wavefields.
In his foundation paper, \cite{Claerbout1971} introduced the imaging condition as the cross-correlation of \textit{``downgoing waves''} generated by the source and \textit{``upgoing waves''} that reach the receivers.
Formula \eqref{eqn:imageP} corresponds to this definition if no reflections occur in the underground.
This assumption is valid when using a one-way wave equation or if the subsurface model 
is very smooth or has a constant impedance. The last option will be considered here for
the pressure-velocity system \eqref{eqn:waveSyst1}--\eqref{eqn:waveSyst2}, which is a two-way wave model that accurately simulates waves propagating in all directions.
Therefore, the resulting imaging procedure does not suffer from dip limitations \citep{Mcmechan1983}, in contrast to one-way wave-equation models.
However, a well-known disadvantage of the wave equation is the correlation of forward and time-reversed wavefields
following the same path, generating long-wave artefacts in the image. Typical examples are diving waves and strong refractions.
One strategy to overcome this difficulty consists in decomposing the source and receivers wavefields into upgoing and downgoing components and only keeping the interesting components in the imaging condition \citep[\eg][]{Hu1987,Liu2011}.
In our case, it is easy to perform such a decomposition thanks to the characteristic fields of the first-order wave system \eqref{eqn:waveSyst1}--\eqref{eqn:waveSyst2}.
Recall that, in one dimension with $z$ as the spatial coordinate, the system is equivalent to
\begin{eqnarray}
  \diff{}{t}\left(p+c\rho\:v_z\right)
    + c \diff{}{z}\left(p+c\rho\:v_z\right) &=& 0, \\
  \diff{}{t}\left(p-c\rho\:v_z\right)
    - c \diff{}{z}\left(p-c\rho\:v_z\right) &=& 0,
\end{eqnarray}
where $v_z$ is the vertical velocity.
These equations correspond to one-way wave-equation models, where waves propagate upwards and downwards, respectively.
By analogy with the previous works, we propose the imaging condition
\begin{eqnarray}
  I(\vec{x}) &=& \sum_\text{shots} \int_{0}^{t_f}
    q^-_\sou(\vec{x},t)\;q^+_\rec(\vec{x},t)\;dt,
  \label{eqn:imageQ}
\end{eqnarray}
where $q^+$ and $q^-$ correspond to the upgoing and downgoing characteristic fields, defined as
\begin{eqnarray}
  q^\pm(\vec{x},t) &=& p(\vec{x},t) \pm c\rho\:\vec{e}_z\cdot\vec{v}(\vec{x},t),
  \label{eqn:charact}
\end{eqnarray}
with the vertical unit vector $\vec{e}_z$, pointing upwards.
% ASC: n codes z axis is usually increasing toward center of Earth...
By using characteristic fields, the new imaging condition \eqref{eqn:imageQ} takes into account upgoing information of the source wavefield and downgoing information of the receiver wavefield.
For steep dips, the above can be easily generalized along the lines of \citep{Liu2011} by also considering decompositions in $\vec{e}_x$ and $\vec{e}_y$.
A complete study of imaging conditions based on characteristic variables is out of the scope of this paper, and will be proposed in future work.

Approximating imaging conditions \eqref{eqn:imageP} and \eqref{eqn:imageQ} with classical numerical integration requires the knowledge of both source and receivers solutions at each time-step.
As mentioned in the introduction, this represents a difficulty because these solutions are simulated with opposite time directions.
Following \cite{Dussaud2008}, we perform both source and receivers simulations backward-in-time, together with a step-by-step computation of the imaging condition.
The backward-in-time source simulation is made possible thanks to a preliminary forward-in-time run, where both the boundary values of fields and the final solution are saved to reproduce the solution in the other (reverse) time direction.

With a nodal DG scheme, we need to save the pressure $p$ and the normal component of the velocity $\vec{n}\cdot\vec{u}$ (called traces of fields in the remainder) only at the boundary nodes, even with high-degree basis functions.
By contrast, when using this strategy with a high-order finite difference scheme, a thick enough halo of data must be saved to compute the thick stencil of the scheme at the boundary \citep[see \eg][]{Yang2014}.
In three dimensions, the amount of data at boundaries increases linearly according to the stencil radius for finite difference schemes whereas the halo zone for a nodal DG scheme is just one node deep.
The thin halo inherent to the nodal DG method then allows for a storage requirement lower than with the widely used high-order finite different schemes.

%% file: schemedg.tex
\subsection{Nodal Discontinuous Galerkin Scheme}
\label{sec:schemeDG}

For each problem, the approximate fields are build on a spatial mesh of the computational domain $\Omega \subset \mathbb{R}^3$, made of non-overlapping tetrahedral cells,
\begin{eqnarray*}
  \Omega &=& \bigcup_{k=1,\hdots,K} \D_k,
\end{eqnarray*}
where $K$ is the number of cells and $\D_k$ is the $k^\text{th}$ cell.
With the nodal discontinuous Galerkin method, the pressure field and each Cartesian component of the velocity field are approached by piecewise polynomial functions.
The discrete unknowns correspond to the values of fields at nodes of elements \citep{Hesthaven2002,Hesthaven2007}.
Over each cell $\D_k$, the approximate fields are then equal to
\begin{eqnarray}
  \disc{p}_k(\vec{x},t) &=& \sum_{n=1}^{N_p} \disc{p}_{k,n}(t)\;\ell_{k,n}(\vec{x}),
  \quad\forall\vec{x}\in \D_k, \label{eqn:appFieldP} \\
  \disc{v}_{i,k}(\vec{x},t) &=& \sum_{n=1}^{N_p} \disc{v}_{i,k,n}(t)\;\ell_{k,n}(\vec{x}),
  \quad\forall\vec{x}\in \D_k,
  \quad i=1,2,3, \label{eqn:appFieldV}
\end{eqnarray}
where $N_p$ is the number of nodes per element, $\disc{p}_{k,n}(t)$ and $\disc{v}_{i,k,n}(t)$ are the values of fields at node $n$ of element $k$, and $\ell_{k,n}(\vec{x})$ is the associated multivariate Lagrange polynomial function.
Denoting the position of the node with $\vec{x}_{k,n}$, we have
\begin{eqnarray*}
  \ell_{k,n}(\vec{x}_{k,m}) &=& \left\{\begin{array}{ll}
    \!\!1, & \text{if } m=n, \\
    \!\!0, & \text{if } m\neq n,
  \end{array}\right.
\end{eqnarray*}
for all $m,n = 1,\hdots,N_p$. In this work, the position of nodes is defined using the warp-and-blend technique \citep{Warburton2006}.
The number of nodes for each element is given by $N_p = (N+1)(N+2)(N+3)/6$, where $N$ is the maximal order of polynomial functions. 

The spatial discretization is obtained from the variational formulation of the problem. For each cell $\D_k$, we are looking for $p(\vec{x},t)\in L_2(\D_k,\mathbb{R}^+)$ and $\vec{v}(\vec{x},t)\in {\bf L}_2(\D_k,\mathbb{R}^+)$ such that
\begin{eqnarray}
  \int_{\D_k}
    \left(\frac{1}{\rho c^2} \diff{p}{t} + \nabla\cdot\vec{v}\right)
    \psi(\vec{x})\:d\vec{x}
  + \int_{\partial \D_k}
    \frac{1}{2} \left(\vec{n}\cdot\jump{\vec{v}}
    - \tau_p \jump{p}\right)\:\psi(\vec{x})\:d\vec{x}
  &=& 0
  \label{eqn:weakFormP} \\
  \int_{\D_k}
    \left(\rho \diff{\vec{v}}{t} + \nabla p\right)
    \cdot\boldsymbol{\psi}(\vec{x})\:d\vec{x}
  + \int_{\partial \D_k}
    \frac{1}{2} \vec{n}
    \left(\jump{p} - \tau_{\vec{v}} \vec{n}\cdot\jump{\vec{v}}\right)
    \cdot\boldsymbol{\psi}(\vec{x})\:d\vec{x}
  &=& 0, \label{eqn:weakFormV}
\end{eqnarray}
for all test functions $\psi(\vec{x})\in L_2(\D^k)$ and $\boldsymbol{\psi}(\vec{x})\in {\bf L}_2(\D^k)$.
In these equations, $\vec{n}$ is the outward unit normal to the cell boundary $\partial \D_k$, and $\jump{p}$ and $\jump{\vec{v}}$ are the jump of fields at this boundary, \ie
\begin{eqnarray*}
  \jump{p} &=& p^+ - p^-, \\
  \jump{\vec{v}} &=& \vec{v}^+ - \vec{v}^-,
\end{eqnarray*}
where the plus and minus subscripts denote nodal values on the side of the neighboring and current cells, respectively.
The formulation is stable if the penalty parameters $\tau_p$ and $\tau_\vec{v}$ are non-negative \citep[see][for a general proof]{Warburton2013}.
With a homogeneous medium, the values $\tau_p=1/(\rho c)$ and $\tau_\vec{v}=\rho c$ correspond to classical upwind fluxes.

The semi-discrete equations are obtained by injecting the approximate representations of fields \eqref{eqn:appFieldP}--\eqref{eqn:appFieldV} in equations \eqref{eqn:weakFormP}--\eqref{eqn:weakFormV}, and using the Lagrange polynomial functions as test functions.
For each element $k$, a system of equations can then be written as
\begin{eqnarray}
  \ddiff{\vec{q}_k^p}{t}
    &=& - \rho_k c_k^2\:\sum_{i=1}^{3} \sum_{j=1}^3
      g_{k,i,j}^{\text{vol}} \vec{D}_{j} \vec{q}_k^{v_i}
    - \sum_{f=1}^{N_f}
      g_{k,f}^{\text{sur}} \mat{L}_{f} \vec{p}_{k,f}^p,
  \label{eqn:dgLocalSchemeP} \\
  \ddiff{\vec{q}_k^{v_i}}{t}
    &=& - \frac{1}{\rho_k} \sum_{j=1}^3
      g_{k,i,j}^{\text{vol}} \vec{D}_{j} \vec{q}_k^p
    - \sum_{f=1}^{N_f}
      g_{k,f}^{\text{sur}} \mat{L}_{f} \vec{p}_{k,f}^{v_i},
  \label{eqn:dgLocalSchemeU}
\end{eqnarray}
where the vectors $\vec{q}_{k}^p$ and $\vec{q}_{k}^{v_j}$ contain the discrete unknowns associated to each field for all nodes of element $k$, and the vectors $\vec{p}_{k,f}^p$ and $\vec{p}_{k,f}^{v_j}$ contain the penalty terms for all face nodes of face $f$.
The first terms of the right-hand sides of these equations correspond to the volume integrals with spatial differential operators of the variational form, while the second terms correspond to the surface integrals.
The matrices $\mat{D}_{j}$ and $\mat{L}_{f}$ are respectively the differentiation matrices and the lifting matrices of the reference element.
The geometric factors $g_{k,i,j}^{\text{vol}}$ and $g_{k,f}^{\text{sur}}$ depend on the shape of each element.
The definitions of the matrices and the factors, and a complete derivation of the semi-discrete equations are given in appendix \ref{appendixScheme.sec}.

A point source term is incorporated into the formulation by adding
\begin{eqnarray}
  \int_{\D_k} f(t)\:\tilde{\delta}(\vec{x}-\vec{x}_0)\:
  \psi(\vec{x})\:d\vec{x},
  \label{eqn:sourceTerm}
\end{eqnarray}
to the right-hand-side of equation \eqref{eqn:weakFormP}. A corresponding term will enter equation \eqref{eqn:dgLocalSchemeP}.
In the term \eqref{eqn:sourceTerm}, the Dirac delta function is replaced by its L2-projection on the polynomial basis,  $\tilde{\delta}(\vec{x}-\vec{x}_0) = \sum_n w_n\ell_{k,n}(\vec{x})$, where the coefficients $w_n$ are obtained by solving the system
\begin{eqnarray*}
  \int_{D_k} \ell_{k,m}(\vec{x})
    \left[\delta(\vec{x}-\vec{x}_0)
    - \sum_{n=1}^{N_p} w_n\ell_{k,n}(\vec{x})\right]d\vec{x}
    &=& 0, \quad\quad \text{for } m=1,\dots,N_p.
\end{eqnarray*}

For sake of clarity, we finally rewrite the semi-discrete equations as the abstract system
\begin{eqnarray}
  \ddiff{\vec{q}_{k}}{t} &=& \vec{r}_{k},
  \label{eqn:dgLocalScheme}
\end{eqnarray}
where $\vec{q}_{k}$ is the vector of all discrete unknowns for element $k$, and $\vec{r}_{k}$ combines the rights-hand side terms of equations \eqref{eqn:dgLocalSchemeP}--\eqref{eqn:dgLocalSchemeU}. The global semi-discrete scheme is simply built by combining the local systems of all elements.

%% file: schemetime.tex
\subsection{Multirate Time-Stepping Scheme}
\label{sec:schemeTime}

The resolution in time is performed by using an explicit multi-rate scheme based on the third-order Adams-Bashforth method.
Such schemes are advantageous in comparison to single-rate ones for problems with refined meshes or involving multiscale dynamics.
The stability of explicit single-rate schemes is guaranteed by the global CFL condition
\begin{eqnarray*}
  \Delta t &\le& C \min_{k}\left\{\frac{h_k}{c_k}\right\},
\end{eqnarray*}
where $C$ is a constant that depends on the time integration scheme and the spatial approximation, $h_k$ is the characteristic size of element $k$, and $c_k$ is the corresponding phase velocity.
The maximum stable time-step $\Delta t$ is thus determined by the element having the smallest ratio $h_k/c_k$.
With multi-rate schemes, the update of approximate fields on each element is performed with a time-step $\Delta t_k$ that is local to the element, and the stability condition becomes local,
\begin{eqnarray}
  \Delta t_k &\le& C \frac{h_k}{c_k}.
  \label{eqn:localCFLCond}
\end{eqnarray}
Larger local time-steps are then allowed for elements having a large ratio $h_k/c_k$, thus providing a reduction of the computational cost of the global scheme.

The multi-rate Adams-Bashforth (MRAB) scheme is efficiently implemented by grouping the elements into levels with fixed time-steps \citep[\eg]{Godel2010,Gandham2015}.
For each element $k$ of level $l$, the local time-step is given by
\begin{eqnarray*}
  \Delta t_{l} &=& 2^{N_\text{levels}-l} \Delta t_\text{local},
\end{eqnarray*}
where $N_\text{levels}$ is the number of levels, and $\Delta t_\text{local}$ is the smallest considered time-step.
The level of each element depends on the maximum local time-step allowed by condition \eqref{eqn:localCFLCond}.
After a first lumping of elements based on this condition, some of them are moved from coarse levels (with larger time-steps) to finer levels (with small time-steps) in order to have no more than one level of difference between two neighboring elements of the mesh.
Therefore, the local time-steps of two neighboring elements are the same or differ with a factor $2$.
Let us denote $\Delta t_\text{global}$ the global time-step, which corresponds to the local time-step of the coarsest level (\ie the level $1$).
For each level $l$, $N_l = 2^{l-1}$ local iterations are needed to achieve a global iteration.

At each local iteration, the discrete unknowns of elements are updated according to the classical third-order Adams-Bashforth formula,
\begin{eqnarray}
    \vec{q}_k^{n+{\color{black}m/N_l}}
  &=& \vec{q}_k^{n+{\color{black}(m-1)/N_l}}
  + {\color{black}\Delta t_l}
    \sum_{s=1}^3 a_s\:\vec{r}_k^{n+{\color{black}(m-s)/N_l}},
  \label{eqn:updateAB}
\end{eqnarray}
with $a_1 = 23/12$, $a_2 = -16/12$ and $a_3 = 5/12$, and where the indices $n$ and $m$ correspond to the global and local time-steppings, respectively. This formula requires the knowledge of the right-hand side vector $\vec{r}_k$ at the three previous local steps. For each step, this vector is thus computed with the values of unknowns local to the element, as well as those of the neighboring element at each interface, according to equations \eqref{eqn:dgLocalSchemeP}--\eqref{eqn:dgLocalSchemeU}. However, at the interface between two elements belonging to two different levels, interface unknowns of the coarse-level element are available at only every two local steps of the fine-level element. For the intermediate step, the unknowns of the coarse-level element evaluated according to the modified Adams-Bashforth formula
\begin{eqnarray}
    \vec{q}_k^{n+{\color{black}(m-\sfrac{1}{2})/N_l}}
  &=& \vec{q}_k^{n+{\color{black}(m-1)/N_l}}
  + {\color{black}\Delta t_l}
    \sum_{s=1}^3 b_f\:\vec{r}_k^{n+{\color{black}(m-s)/N_l}},
  \label{eqn:updateABcoarse}
\end{eqnarray}
with the coefficients $b_1 = 17/24$, $b_2 = -7/24$ and $b_3 = 2/24$.

This procedure is straightforwardly adapted for backward-in-time simulations.
It suffices to consider that time decreases when the indexes $n$ and $m$ increase.
No changes are then required in the update schemes, but the penalty parameters $\tau_p$ and $\tau_\vec{v}$ now must be strictly negative to stabilize this backward-in-time scheme.

%% file: schemertm.tex
 \subsection{Evaluation of Imaging Condition}
 \label{sec:schemeRTM}

We will now focus on the computation of the image $I(\vec{x})$, defined by equation \eqref{eqn:imageP}.
The generalization to the imaging condition \eqref{eqn:imageQ} is straightforward.
For one sample of source-receivers signals, the time integral of $I(\vec{x})$ is evaluated
time-step by time-step along a backward-in-time computation of both source solution $p_\sou(\vec{x},t)$ and receivers solution $p_\rec(\vec{x},t)$, as explained in section \ref{inverseProblem.sec}.
Each local time-step contribution to the time integral,
\begin{eqnarray}
  \int_{t^{n+(m-1)/N_l}}^{t^{n+m/N_l}}
    p_\sou(\vec{x},t)\;
    p_\rec(\vec{x},t)\:dt,
  \label{eqn:crossCorrContLoc}
\end{eqnarray}
with $t^{n+m/N_l}=(n+m/N_l)\dt$, is computed at each node $\vec{x}$ and the values are accumulated over the steps to provide $I(\vec{x})$.

Integral \eqref{eqn:crossCorrContLoc} could typically be evaluated using the trapezoidal rule
\begin{eqnarray}
  \frac{\dt_l}{2}
  \left(p_\sou^{n+m/N_l}(\vec{x})\;
        p_\rec^{n+m/N_l}(\vec{x})
      - p_\sou^{n+(m-1)/N_l}(\vec{x})\;
        p_\rec^{n+(m-1)/N_l}(\vec{x})\right),
  \label{eqn:trapRule}
\end{eqnarray}
which is a first-order approximation in time to the exact integral.
We propose a better evaluation considering that the fields are obtained with a Adams-Bashforth scheme.
Let us recall that this scheme uses on polynomial representations of fields
\begin{eqnarray}
p_\sou(\vec{x},t) &=& p_\sou^{n+(m-1)/N_l}(\vec{x})
  + \sum_{s=1}^3 \left(\int_{0}^{t^\star(t)} \ell_{s}(t')\:dt'\right)
    r_\sou^{n+(m-s)/N_l}(\vec{x}), \label{eqn:ABrepS} \\
p_\rec(\vec{x},t) &=& p_\rec^{n+(m-1)/N_l}(\vec{x})
  + \sum_{s=1}^3 \left(\int_{0}^{t^\star(t)} \ell_{s}(t')\:dt'\right)
    r_\rec^{n+(m-s)/N_l}(\vec{x}), \label{eqn:ABrepR}
\end{eqnarray}
with $t^\star(t) = \left(t-t^{n+(m-1)/N_l}\right)/\dt_l$,
where $\ell_1(t)$, $\ell_2(t)$ and $\ell_3(t)$ are Lagrange interpolation functions with interpolation nodes at $0$, $-1$ and $-2$, and $r_\sou^{n+(m-s)/N_l}$ and $r_\rec^{n+(m-s)/N_l}$ are the components of right-hand side vectors corresponding to pressure field.
Injecting representations \eqref{eqn:ABrepS}-\eqref{eqn:ABrepR} in integral \eqref{eqn:crossCorrContLoc} gives
\begin{eqnarray}
   & & \dt_l   \; p_\sou^{n+(m-1)/N_l}(\vec{x})\;p_\rec^{n+(m-1)/N_l}(\vec{x}) \nonumber \\
   &+& \dt_l^2 \: \sum_{s=1}^3 a_s\;p_\sou^{n+(m-1)/N_l}(\vec{x})\;r_\rec^{n+(m-s)/N_l}(\vec{x}) \nonumber \\
   &+& \dt_l^2 \: \sum_{s=1}^3 a_s\;r_\sou^{n+(m-1)/N_l}(\vec{x})\;p_\rec^{n+(m-s)/N_l}(\vec{x}) \nonumber \\
   &+& \dt_l^3 \: \sum_{s_1=1}^3 \sum_{s_2=1}^3 C_{s_1 s_2}\;
      r_\sou^{n+(m-s_1)/N_l}(\vec{x})\;r_\rec^{n+(m-s_2)/N_l}(\vec{x}),
  \label{eqn:updateABcorr}
\end{eqnarray}
with the coefficient matrix
\begin{eqnarray*}
\mat{C} =
\begin{pmatrix}
 \frac{4703}{5040} & -\frac{457}{840}  & \frac{52}{315} \\ 
 -\frac{457}{840}  & \frac{103}{315}   & -\frac{251}{2520} \\ 
 \frac{52}{315}    & -\frac{251}{2520} & \frac{17}{560}
\end{pmatrix}.
\end{eqnarray*}
This formula gives the exact value of integral \eqref{eqn:crossCorrContLoc} for fields that are third-order approximation in time.
Therefore, formula \eqref{eqn:updateABcorr} theoretically keeps accuracy of numerical fields with an $\mathcal{O}(\dt^3)$ numerical error, while approximation \eqref{eqn:trapRule} gives an $\mathcal{O}(\dt)$ error.

It is worthwhile to mention that this numerical evaluation is based on values of fields and right-hand side vectors that are available to update the fields with formula \eqref{eqn:updateAB}.
The contribution \eqref{eqn:updateABcorr} to the imaging condition is thus computed at each step right before the update of fields, without significant supplementary cost in comparison with approximation \eqref{eqn:trapRule}.

%% file: computationocca.tex
\subsection{Multi-Threading Programming with OCCA}
\label{computationOCCA}

OCCA is a unified framework for multi-threading programming that can be used for several shared-memory hardware architectures, such as CPUs, GPUs and Intel's Xeon Phi.
The current version translates a single implementation of computational kernels to the OpenMP, OpenCL, pThreads, Intel COI and CUDA languages.
It has been developed to maintain portability and performance together with platform-choice flexibility.
Using this programming approach therefore allows customized implementations of algorithms for several computing devices with a single code.

The OCCA library provides a host API and a device API. The host API gives tools to generates a self-contained context and command queue from hardware devices, to allocate and manage device memory, and finally to compile and execute kernels.
The device API allows us to write kernels that are compiled at run-time, and that send instructions to the device when executed.
The kernel language of OCCA mirrors those used for GPU programming.
Threads (work-items) are grouped into groups (work-groups), which compose the main grid.
The work-groups are queued for execution onto the available multiprocessors, and work-items are executed in parallel.
The optimum choices of the number of work-groups, and the number of work-items per work-group depend on the algorithm and the characteristics of the device in use.
These number are defined by the user, before the execution of each kernel, thanks to the host API.

Further information about OCCA can be found in a white paper \citep{Medina2014}. The latest developments are available on the website \texttt{http://www.libocca.org/}.

%% file: computationkernels.tex
\subsection{Implementation with One Multi-Threading Device}
\label{computationKernels}

The complete computational procedure for the forward time-stepping is given in algorithm \ref{alg:forwardMRAB}.
It is straightforwardly adapted for the backward simulation.
For each element $k$, $\texttt{q}_k$ stores all the local values of fields ($p$ and each Cartesian component of $\vec{v}$), and $\texttt{qf}_k$ stores a copy of traces corresponding to face nodes ($p$ and $\vec{n}\cdot\vec{v}$, where $\vec{n}$ is the outward unit normal to the face).
When updating the right-hand-side vector, $\texttt{q}_k$ is used to compute volume terms, while $\texttt{qf}_k$ is used for the surface terms.
The array $\texttt{rhs}_k$ stores the right-hand side vectors corresponding to the three previous local time-steps.
$n_\text{local}$ and $N_\text{local}$ are the local step index and the number of local steps of the finest level.

\begin{algorithm}[!b]
\caption{Forward time-stepping procedure}
\label{alg:forwardMRAB}
\textbf{initialization} \\
\quad initialize $\texttt{q}_k$ for each element $k$ \\
\quad initialize $\texttt{qf}_k$ for each element $k$ \\
\quad initialize $\texttt{rhs}_k$ for each element $k$ \\
\For{$n_{\text{\normalfont global}}=1,2,\dots,N_{\text{\normalfont global}}$}{
\For{$n_{\text{\normalfont local}}=1,2,\dots,N_{\text{\normalfont local}}$}{
$t_{\text{\normalfont start}}=[(n_{\text{\normalfont global}}-1) N_{\text{\normalfont local}} +(n_{\text{\normalfont local}}-1)]\cdot\Delta t_\text{local}$ \\
$t_{\text{\normalfont new}}=t_{\text{\normalfont start}}+\Delta t_\text{local}$ \\
\For{\normalfont each element $k$ with unknowns evaluated at $t_{\text{\normalfont start}}$}{
update $\texttt{rhs}_k$ for $t=t_{\text{\normalfont start}}$ \\
}
\For{$l=N_\text{\normalfont levels},...,2,1$}{
\If{\normalfont the unknowns of level $l$ must be evaluated at $t_\text{new}$}{
\For{\normalfont each element $k$ of level $l$}{
update $\texttt{q}_k$ and $\texttt{qf}_k$
for $t=t_{\text{\normalfont new}}$
with update scheme \eqref{eqn:updateAB}
}
}
}
\vspace{0.1cm}
\For{\normalfont each element $k$ that has not been updated at this local time-step, \\
but that is coarse-level neighbour of evaluated elements}{
\vspace{0.1cm}
update $\texttt{qf}_k$
for $t=t_{\text{\normalfont new}}$
with update scheme \eqref{eqn:updateABcoarse}
}
}
}
\end{algorithm}

This global procedure can be naturally decomposed into several computational kernels.
These computational kernels are implemented in separate OCCA kernels, which allows us to optimize each task considering the characteristics of both the task and the device.
Thereby, the loop iterations of algorithms \ref{alg:forwardMRAB} are performed by the main process (on the `host'), which calls the different kernels to execute operations on the device.
Our implementation has three main kernels, which are called at each local time-step in this order:
\begin{enumerate}
\setlength\itemsep{0em}
\item The \textit{volume kernel} computes the volume terms and stores the result into array $\texttt{rhs}_k$.
\item The \textit{surface kernel} computes the surface terms and updates $\texttt{rhs}_k$ with the result.
\item The \textit{update kernel} performs the local time-stepping with Adams-Bashforth schemes \eqref{eqn:updateAB} and \eqref{eqn:updateABcoarse}.
\end{enumerate}
A supplementary kernel (the \textit{source kernel}) generates signals at sources and receivers by adding a specific term in the right-hand side vector.
For the RTM procedure, a specific kernel (the \textit{image kernel}) updates an image array $\texttt{I}_k$ with the contribution of the current local time-step to the imaging condition.

In implementing these kernels, we seek to optimize the use of the computational units of the hardware device and minimize waiting time when transferring data, in order to optimize run time.
These questions have been studied by \cite{Klockner2009,Klockner2012,Klockner2013} for the implementation of nodal discontinuous Galerkin schemes on GPU.
The ideas have then been applied with an Adams-Bashforth multi-rate time-stepping by \cite{Godel2010} and \cite{Gandham2015} in different wave contexts.

\subsubsection*{Memory on the device}

The locality of memory storage is addressed by storing all the data required by the numerical schemes on the device (\eg in the global memory of a GPU).
A floating-point array \texttt{q} stores all the discrete unknowns, and the array \texttt{qf} contains a copy of traces associated to face nodes.
The array \texttt{rhs} is reserved for the right-hand side vectors associated to the three previous time-steps.
In a RTM procedure, all these arrays are duplicated for source and receivers simulations, and the imaging condition is stored in the specific array \texttt{I}.
When using an accelerator, such as a GPU, these arrays are only stored in its global memory and are never transferred to the RAM of the compute node, at any moment, except to export the final RTM image.

During an initialization procedure, the differentiation matrices $\mat{D}_j$ and the lifting matrices $\mat{L}_f$ are precomputed and stored on the device in the aggregated arrays \texttt{Drst} and \texttt{Lift}.
The Jacobian matrices $\partial\boldsymbol{\Psi}_k/\partial\vec{r}$ of all elements are stored in \texttt{vGeoFac}, and both Cartesian the components of normal vector and the Jacobian ratios $J_{k,f}/J_f$ are stored in \texttt{sGeoFac}.
The physical parameters $\rho_k$ and $c_k$ are managed using a database principle: the array \texttt{phyDB} contains the parameters values for each kind of medium, and the mapping between medium indexes and mesh cells is stored in the array \texttt{phyMap}.
Finally, several arrays list elements associated to each multi-rate time-stepping level, and to the coarse neighbours for the MRAB strategy.

\begin{table}[!b]
\centering
\begin{tabular}{l|l|l} \hline
Symbol & Dimensions & Definition \\ \hline
\texttt{q} & $K \cdot N_{\fields} \cdot N_{\pad}$ & Values of fields at nodes \\
\texttt{qf} & $K \cdot N_{\traces} \cdot N_{f} \cdot N_{fp} $ & Values of traces at face nodes \\
\texttt{rhs} & $3 \cdot K \cdot N_{\fields} \cdot N_{\pad}$ & Right-hand side terms \\
\texttt{I} & $K \cdot N_{\pad}$ & Imaging condition \\
\texttt{Drst} & $N_p^2 \cdot N_{\dim}$ & Differentiation matrices \\
\texttt{Lift} & $N_{f} \cdot N_{fp} \cdot N_p$ & Lifting matrices \\
\texttt{vGeoFac} & $K \cdot N_{\dim}^2$ & Geometric factors for volume terms \\
\texttt{sGeoFac} & $K \cdot 4 \cdot N_f$ & Geometric factors for surface terms \\ \hline
\end{tabular}
\caption{List of main arrays stored in the global memory of the device.
Dimensions are given from the coarsest to the finest granularity of storage.}
\label{tab:arrays}
\end{table}

In order to take full advantage of the cache memory, the different nodal values corresponding to a same field and a same element are stored contiguously in the array \texttt{q}.
When fetching values from the memory to device registers, the memory bus operates by blocks of data instead of individual items.
Since these nodal values must be used together to compute the first matrix-vector products of the right-hand sides of equations \eqref{eqn:dgLocalSchemeP}-\eqref{eqn:dgLocalSchemeU}, a contiguous storage is advantageous.
We improve the utilization of the memory bus by padding the blocks of nodal values in \texttt{q}.
$N_{\pad}$ array elements are thus reserved instead of $N_p$ for each block, where $N_{\pad}\ge N_p$ is chosen considering the characteristics of the bus.
Then, the blocks of $N_{\pad}$ array elements corresponding to the different fields for a same element are stored contiguously. The coarsest granularity of storage thus corresponds to the element indexes.
A similar granularity is used for \texttt{qf}, \texttt{rhs} and \texttt{I}.
The dimensions and granularity of the main arrays are given in table \ref{tab:arrays}.

\subsubsection*{Kernels}

The volume and surface kernels, shown in algorithms \ref{alg:volumeKernel} and \ref{alg:surfaceKernel}, are conceived and optimized in a similar way.
When one of them is executed, the volume or surface terms are computed and stored in \texttt{rhs} for all nodes that must be updated at the current local time-step.
For both kernels, each thread (work-item) performs the operations for the $N_\text{fields}$ items of array \texttt{rhs} associated to a given node and the current time-step.
Each work-group processes the nodes of several elements, $K_\blkV$ for the volume kernel and $K_\blkS$ for the surface kernel.
These parameters provide a way to tune the occupation of the device for each kernel, and then to adjust the load balancing.
The volume kernel is then compiled for $\text{ceil}(K/K_\blkV)$ work-groups of $K_\blkV \cdot N_p$ work-items, and the surface kernel is compiled for $\text{ceil}(K/K_\blkS)$ work-groups of $K_\blkS \cdot \max(N_p, N_f N_{fp})$ work-items.

\input{kernelvolume}

\input{kernelsurface}

The computation of volume and surface terms are efficiently processed in three sub-tasks, which correspond to the three OCCA inner loops of algorithms \ref{alg:volumeKernel} and \ref{alg:surfaceKernel}.
First, the geometric factors and the physical parameters, which are shared by all the nodes and face nodes of a given element, are transported from the global device memory to shared arrays.
The second sub-task consists in building the vectors that are used in the matrix-vector products of equations \eqref{eqn:dgLocalSchemeP}--\eqref{eqn:dgLocalSchemeU}.
For the volume kernel, each work-item fetches the fields values of its assigned node, computes the linear combinations of velocity unknowns, and stores the results in shared arrays.
In addition, the right-hand-side vectors of the two previous time-steps are updated.
For the surface kernel, each work-item fetches the field values of its assigned face node, both for the current and neighbouring element.
Then, it computes the penalty terms, multiplies them with the ratios of Jacobians, and stores the results in shared arrays.
Finally, the matrix-vector products are computed in the third sub-tasks of kernels, and the results are added to \texttt{rhs}.
Each work-item effectively performs only one scalar product of vectors for each matrix-vector product.
This operation is optimized thanks to the coherent granularity of storage and some supplementary fine-tuning (\eg loop unrolls and storage of intermediates parameters in constant memories).
Further details about these strategies can be found in the works of \cite{Klockner2009,Klockner2012,Klockner2013}.

\input{kernelupdate}

The update kernel performs the time-stepping \eqref{eqn:updateAB} for all nodal values at a given time level $l$, stores the new values in arrays \texttt{q} and updates the traces in \texttt{qf}, as shown in algorithm \ref{alg:updateKernel}.
Each work-item updates the nodal values associated to a given node, and each work-group deals with $K_\blkS$ elements.
The operations are again achieved in sub-tasks. 
First, the values are updated and stored in a shared array.
After a synchronization, they are transported into arrays \texttt{q} and \texttt{qf}, requiring respectively $N_p$ and $N_f N_{fp}$ work-items for the most inner loop.
The size of the work-group is thus $K_\blkS \cdot \max(N_p, N_f N_{fp})$.
This kernel is also used to update coarse neighbour elements with the time-stepping \eqref{eqn:updateABcoarse}.
The image kernel is completely analogous.

%% file: kernelvolume.tex
\begin{algorithm}[!t]
\caption{Volume kernel}
\label{alg:volumeKernel}
\textbf{input} \\
\quad pointers $^*\texttt{q}$, $^*\texttt{rhs}$ \\
\quad pointers $^*\texttt{Drst}$, $^*\texttt{vGeoFac}$, $^*\texttt{phyMap}$, $^*\texttt{phyDB}$ \\

\For{\normalfont each block \texttt{b} of elements}{
shared arrays \texttt{P}, \texttt{UdotGr}, \texttt{UdotGs}, \texttt{UdotGt} \ {\color{gray}(arrays $K_\blkV \cdot N_p$)} \\
shared arrays \texttt{vGeoFacWG} \ {\color{gray}(array $K_\blkV \cdot 9$)} \\
shared arrays \texttt{phyParWG} \ {\color{gray}(array $K_\blkV \cdot 2$)} \\

\For{\normalfont each element \texttt{k} of block \texttt{b}}{
\For{\normalfont each node \texttt{n} of element \texttt{k}}{
    load the geometric factors in \texttt{vGeoFacWG} for element \texttt{k} \\
    load the physical parameters in \texttt{phyParWG} for element \texttt{k} \\
}
}

\For{\normalfont each element \texttt{k} of block \texttt{b}}{
\For{\normalfont each node \texttt{n} of element \texttt{k}}{
    read the values of fields from \texttt{q} \\
    compute the linear combinations for the $p$ equation \\
    store \texttt{P}, \texttt{UdotGr}, \texttt{UdotGs} and \texttt{UdotGt}
}
}

\For{\normalfont each element \texttt{k} of block \texttt{b}}{
\For{\normalfont each node \texttt{n} of element \texttt{k}}{
    read values of the differentiation matrices from \texttt{Drst} \\
    compute $\texttt{Dr}\cdot\texttt{UdotGr}
        +\texttt{Ds}\cdot\texttt{UdotGs}
        +\texttt{Dt}\cdot\texttt{UdotGt}$ for the $p$ equation \\
    compute $\texttt{Dr}\cdot\texttt{P}$, $\texttt{Ds}\cdot\texttt{P}$
        and $\texttt{Dt}\cdot\texttt{P}$, and the linear combinations for the $v_i$'s equations \\
    compute the volume terms and store in \texttt{rhs}
}
}

}
\end{algorithm}

%% file: kernelsurface.tex
\begin{algorithm}[!t]
\caption{Surface kernel}
\label{alg:surfaceKernel}
\textbf{input} \\
\quad pointers $^*\texttt{qf}$, $^*\texttt{rhs}$ \\
\quad pointers $^*\texttt{Lift}$, $^*\texttt{sGeoFac}$, $^*\texttt{phyMap}$, $^*\texttt{phyDB}$, $^*\texttt{typeMap}$ \\

\For{\normalfont each block \texttt{b} of elements}{

shared arrays \texttt{pPena}, \texttt{uPena}, \texttt{vPena}, \texttt{wPena} \ {\color{gray}(arrays $K_\blkS \cdot N_f \cdot N_{fp}$)} \\
shared arrays \texttt{sGeoFacWG} \ {\color{gray}(array $K_\blkS \cdot 4 \cdot N_f$)} \\
shared arrays \texttt{phyParWG} \ {\color{gray}(array $K_\blkS \cdot 2$)} \\

\For{\normalfont each element \texttt{k} of block \texttt{b}}{
\For{\normalfont each node \texttt{n} of element \texttt{k}}{
    load the geometric factors in \texttt{sGeoFacWG} for element \texttt{k} \\
    load the physical parameters in \texttt{phyParWG} for element \texttt{k}
}
}

\For{\normalfont each element \texttt{k} of block \texttt{b}}{
\For{\normalfont each face node \texttt{nf} of element \texttt{k}}{
    read the interior values of fields from \texttt{qf} \\
    read the type of face from \texttt{typeMap} \\
    \If{\normalfont face at domain boundary}{
      define the exterior values of fields with the condition \\
    }
    \ElseIf{\normalfont face at interface}{
      read the exterior values of fields from \texttt{qf} \\
    }
    compute the  components of penalty vectors \\
    multiply with the ratios of Jacobians \\
    store in \texttt{pPena}, \texttt{uPena}, \texttt{vPena} and \texttt{wPena}
}
}

\For{\normalfont each element \texttt{k} of block \texttt{b}}{
\For{\normalfont each node \texttt{n} of element \texttt{k}}{
    read the values of lifting matrices from \texttt{Lift} \\
    compute
      $\texttt{Lift}\cdot\texttt{pPena}$,
      $\texttt{Lift}\cdot\texttt{uPena}$,
      $\texttt{Lift}\cdot\texttt{vPena}$ and
      $\texttt{Lift}\cdot\texttt{wPena}$ \\
    add the resulting surface terms to \texttt{rhs}
}
}

}
\end{algorithm}

%% file: kernelupdate.tex
\begin{algorithm}[!t]
\caption{Update kernel}
\label{alg:updateKernel}
\textbf{input} \\
\quad pointers $^*\texttt{q}$, $^*\texttt{qf}$, $^*\texttt{rhs}$ and $^*\texttt{Fmask}$ \\

\For{\normalfont each block \texttt{b} of elements}{

shared array \texttt{qNew} \ {\color{gray}(array $N_\fields \cdot K_\blkS \cdot N_p$)} \\

\For{\normalfont each element \texttt{k} of block \texttt{b}}{
\For{\normalfont each node \texttt{n} of element \texttt{k}}{
    compute the updated fields and store the values in $\texttt{qNew}$
}
}

\For{\normalfont each element \texttt{k} of block \texttt{b}}{
\For{\normalfont each node \texttt{n} of element \texttt{k}}{
    update \texttt{q} with the new values stored in \texttt{qnew}
}
}

\For{\normalfont each element \texttt{k} of block \texttt{b}}{
\For{\normalfont each face node \texttt{nf} of element \texttt{k}}{
    read node index \texttt{n} corresponding to face node \texttt{nf} in \texttt{Fmask} \\
    update \texttt{qf} with the new values stored in \texttt{qnew}
}
}

}
\end{algorithm}

%% file: computationmpi.tex
\subsection{Towards an Implementation for Cluster of Devices}
\label{computationMPI}

The main challenge for large-scale computations with a large number of cores (with or without accelerators) is to reach a good parallel scalability for the final implementation.
This requires to take care of two critical aspects:  load balancing and managing the latency inherent to data transfers between compute nodes.

In our implementation, the device associated to each compute node of the cluster performs the whole computation for a sub-domain corresponding to one part of the mesh.
The procedure presented in the previous section is naturally applied to each device.
At each local time-step, MPI communications transfer the traces of fields situated at the interface between two sub-domains when necessary.
For the backward run of the RTM procedure, both source and receivers simulations are performed using the same mesh partition: this ensures the computation of the imaging condition without extra data transfers between processes.

The mesh partition is generated using METIS with a weighting strategy.
The weight $N_l = 2^{l-1}$ is associated to each cell, where $l$ is the time level of the cell.
As explained in section \ref{sec:schemeTime}, $N_l$ is also the number of local iterations required to achieve a global iteration for elements of level $l$.
This number then is proportional to the compute load required for the corresponding element, which allows to balance the global load between the compute nodes.

Our experience has shown that the elements belonging to the finest time level were generally isolated in the mesh (see benchmark of section \ref{sec:benchRTM}).
In the multi-rate time-stepping strategy, these elements are updated at every smallest time-step, and only those are updated for half smallest time-steps.
By preventing METIS to place them at interfaces between mesh parts, we remove the need for MPI communications when only those are updated, which represents half of all potential MPI communications.
This is done by lumping each finest element (or group of finest elements) with his neighbors before METIS partition, and by separating them after the procedure.

%% file: computationrtm.tex
\subsection{Memory Transfers for MPI Communications and RTM Procedure}
\label{computationRTM}

We now detail our strategies for the memory transfers that are required to connect sub-domains together, and to record/replay fields at the domain boundary for the source simulation in the RTM procedure.
Both operations require different kinds of memory transfers:
\begin{enumerate}
\item The connection of sub-domains is obviously made through MPI communications (host-host transfers).
Data to send must be pre-fetched from device memory, while received data are loaded on the device memory right after.
Host-device transfers are then required.
\item Boundary data recorded for the source simulation must be stored in a memory large enough.
The device memory usually being too small, the data is accumulated and stored either in the host memory --- more precisely, in its random-access memory (RAM) --- or on a hard disk disk drive (HDD).
We considered both options.
This then requires host-device transfers with optionally RAM-HDD transfers.
\end{enumerate}

\begin{figure}
\centering
\begin{tabular}{c@{\hspace{1.5cm}}c}
  \includegraphics[width=0.36\textwidth,natwidth=7.6cm,natheight=13.8cm]{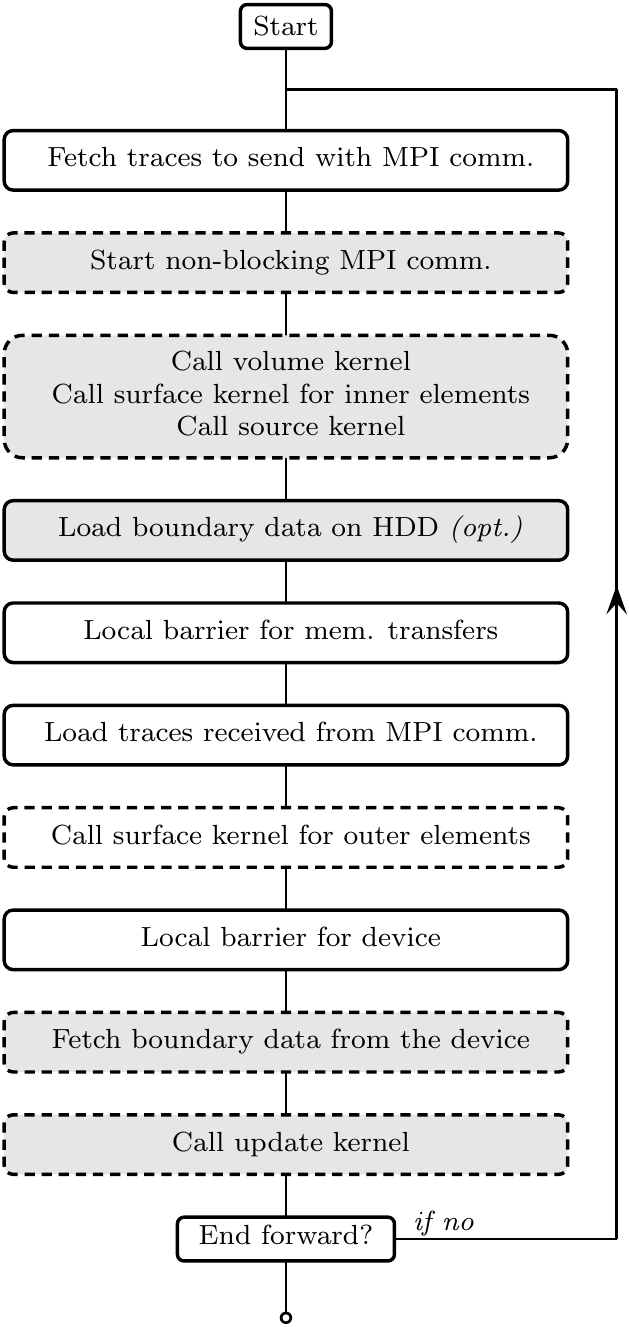} &
  \includegraphics[width=0.36\textwidth,natwidth=7.6cm,natheight=13.8cm]{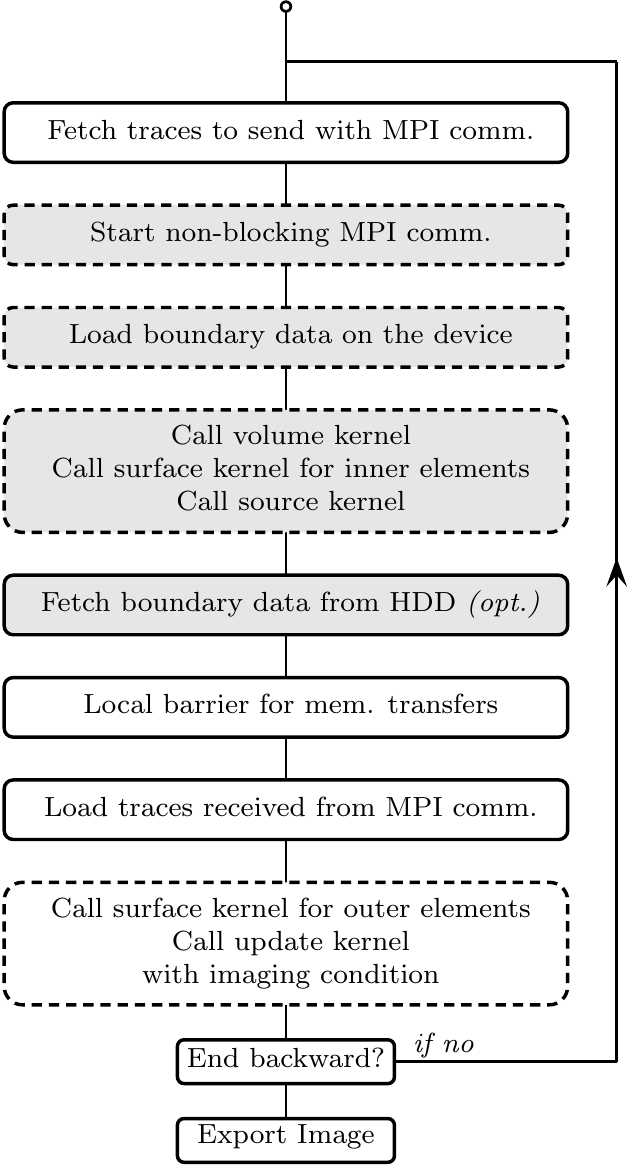}
\end{tabular}
\caption{
Flow chart of the complete RTM procedure with the forward phase (left) and the backward phase (right).
Each iteration of loops corresponds to a local time-step update (forward or backward).
For the backward phase, each kernel performs the same operations for both source and receivers simulations.
Boxes with dashed borders correspond to operations that are non-blocking for the host.
Successive boxes with gray background correspond to independent operations that can be performed at the same time.
}
\label{fig:flowchart}
\end{figure}

The flow chart of the complete RTM procedure with the memory transfers is shown in figure \ref{fig:flowchart}.
Our goal was to hide latency inherent to transfers regardless of the kind of transfer.
Therefore, we used non-blocking transfers whenever it was possible, in order to perform the different transfers alongside computations on the device.
While our implementation runs on clusters with only CPUs, it is mainly conceived for clusters with accelerators.
In that context, calling a kernel is a non-blocking operation for the compute (host) node.
The operations mentioned in figure \ref{fig:flowchart} are performed by the compute node, and data transfers from/to are written from that point of view.
We describe hereafter several of the strategies that were employed.

The majority of data transfers are performed at the beginning of iterations, alongside kernels that do not need to immediately process transfered data: the volume kernel, the surface kernel for inner elements, and the source kernel.
On the flow chart (figure \ref{fig:flowchart}), these operations correspond to the two main groups of successive gray boxes.
By calling twice the surface kernel, once for inner elements (that do not touch the border of the sub-domain) and once for outer elements, we increase the compute time available to hide data transfers.
Before starting MPI communications, data to be sent to neighboring sub-domains is fetched from the device memory to the host memory.
Because this data must already be on the host memory to start communications, blocking host-device transfers are used.
By contrast, during the backward simulation, boundary data can be loaded onto the device memory with non-blocking host-device transfers.
RAM-HDD exchanges are made in a blocking way after all the non-blocking operations, but occur alongside them.
When all these operations are completed, data received with MPI communications is loaded on the device memory, and surface terms of outer elements are computed.
This ends the computation of the right-hand-side terms.

Different containers are used to store and to transfer boundary data during the RTM procedure.
A specific buffer array is allocated on the device memory for temporary storage.
When the surface kernel is executed for outer elements, the traces corresponding to boundary nodes are copied into that buffer during the forward phase, and are read from it during the backward phase.
This explain why, during the forward phase, boundary data are transferred from the device memory to the host memory at the end of each iteration.
In the host memory (\ie the RAM of the compute node), accumulated boundary data is stored in a sequence container with double ended queue.
At each forward iteration, a new container is added at one end of the sequence with the current boundary data.
When using a HDD to unload the RAM, data is transferred to the HDD by reading the sequence container from its other end, with a time delay of one global time-step.
This means that the sequence container always contains boundary data corresponding to one global time-step.
Since both host-device and RAM-HDD transfers involve different pieces of data, they can be performed simultaneously.
The strategy is identical for the backward phase, except that HDD-RAM transfers are performed one global time-step in advance.

%% file: benchvalid.tex
\subsection{Validation Case}

In order to validate our implementation, numerical convergence as well as run times, we use a reference benchmark proposed by \cite{Zhebel2014} and \cite{Minisini2012}.

%%% DESCRIPTION

\begin{figure}
  \centering
  \begin{tabular}{cc}
  (a) Velocity model & (b) Snapshot of pressure \\
  \includegraphics[width=0.45\textwidth,natwidth=931px,natheight=834px]{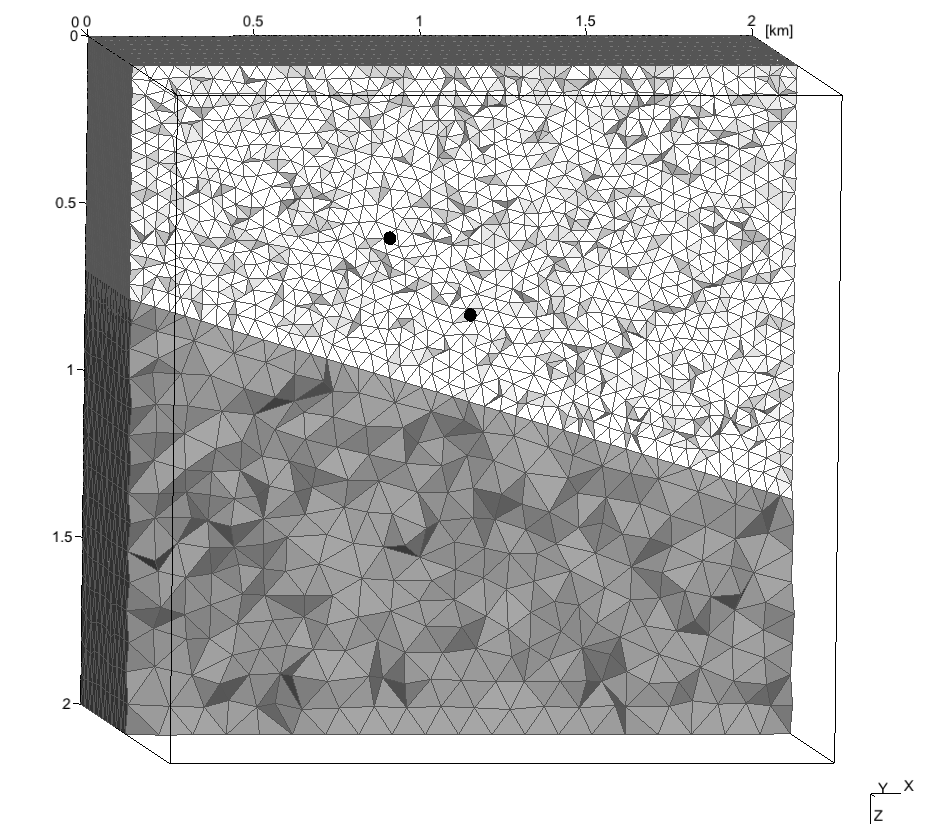}
  \quad &
  \includegraphics[width=0.45\textwidth,natwidth=931px,natheight=834px]{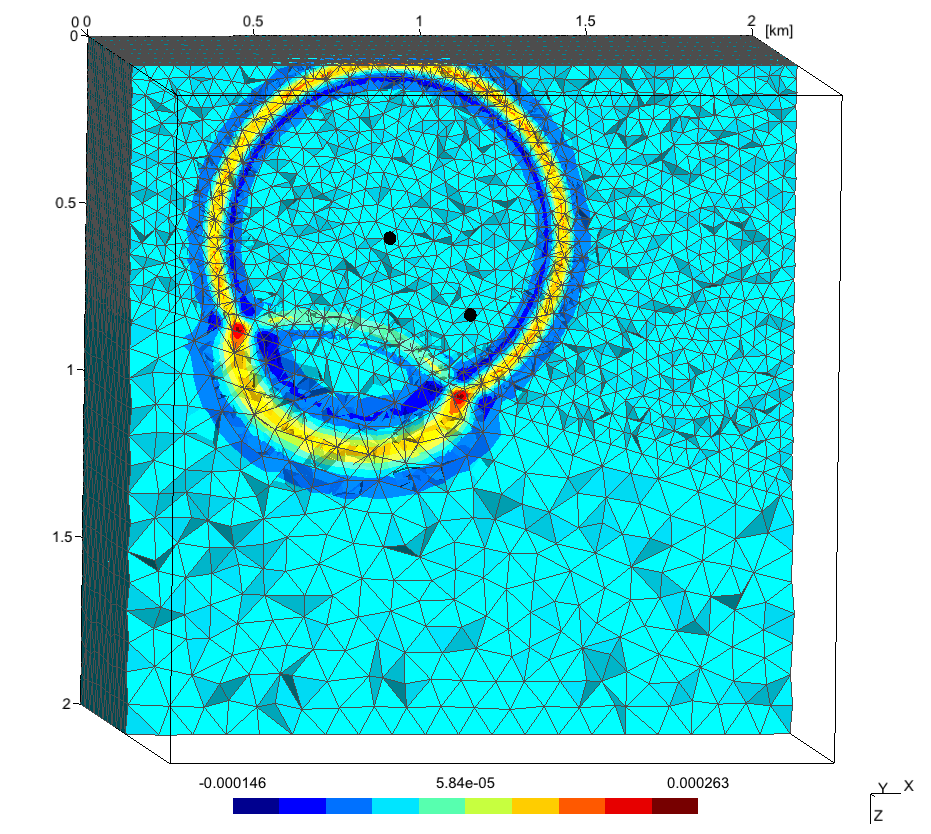}
  \end{tabular}
  \caption{Representation of the half part of the computational domain with the coarsest mesh \textnormal{(a)} and snapshot of the pressure at one instant \textnormal{(b)}.
The velocity is $1.5\:\km/\s$ in the upper medium (light gray) and $3\:\km/\s$ in the lower one (dark gray).
The source position (upper black point) and the receiver position (lower black point) are represented on both figures.
The box indicates the total size of the computational domain.}
  \label{fig:case1:bench}
\end{figure}

The computational domain of the benchmark is a cube $\Omega=[0.2\:\km]\times[0.2\:\km]\times[0.2\:\km]$ made of two media separated with a dipping plane interface (figure \ref{fig:case1:bench}a).
The density is $1\:\g/\cm^3$ in both media, and the velocity is $1.5\:\km/\s$ in the upper one and $3\:\km/\s$ in the lower one.
The plane interface runs from $0.7\:\km$ in depth at $x=0$, to $1.3\:\km$ at $x=2\:\km$.
A Ricker pulse with a peak frequency of $12\:$Hz is generated at a point source located above the interface at position $\vec{x}_s = (779.7\:\m, 1000\:\m, 516.3\:\m)$, and creates a spherical wave (figure \ref{fig:case1:bench}b).
The peak of the Ricker pulse is generated at instant $t_p=0.1127\:\s$ and the final time of the simulation is $t_f=0.8127\:\s$.
During the simulation, the pressure is recorded at a receiver situated at position $\vec{x}_r = (1023.9\:\m, 1000\:\m, 746.2\:\m)$.
The recorded signal is compared to the exact solution over the period $[t_p,t_f]$ using the relative error defined as \citep{Zhebel2014}
\begin{eqnarray}
  \sqrt{ \frac{\int_{t_p}^{t_f} \left|p_\text{num}-p_\text{ref}\right|^2 dt}{\int_{t_p}^{t_f} \left|p_\text{ref}\right|^2 dt} },
\end{eqnarray}
where $p_\text{num}(t)$ and $p_\text{ref}(t)$ are respectively the recorded and reference signals.
Because of the chosen final time, the boundary conditions have no influence on the error.

%%% CONVERGENCE

\begin{figure}[!t]
  \centering
  \includegraphics[width=0.55\textwidth,natwidth=576px,natheight=432px]{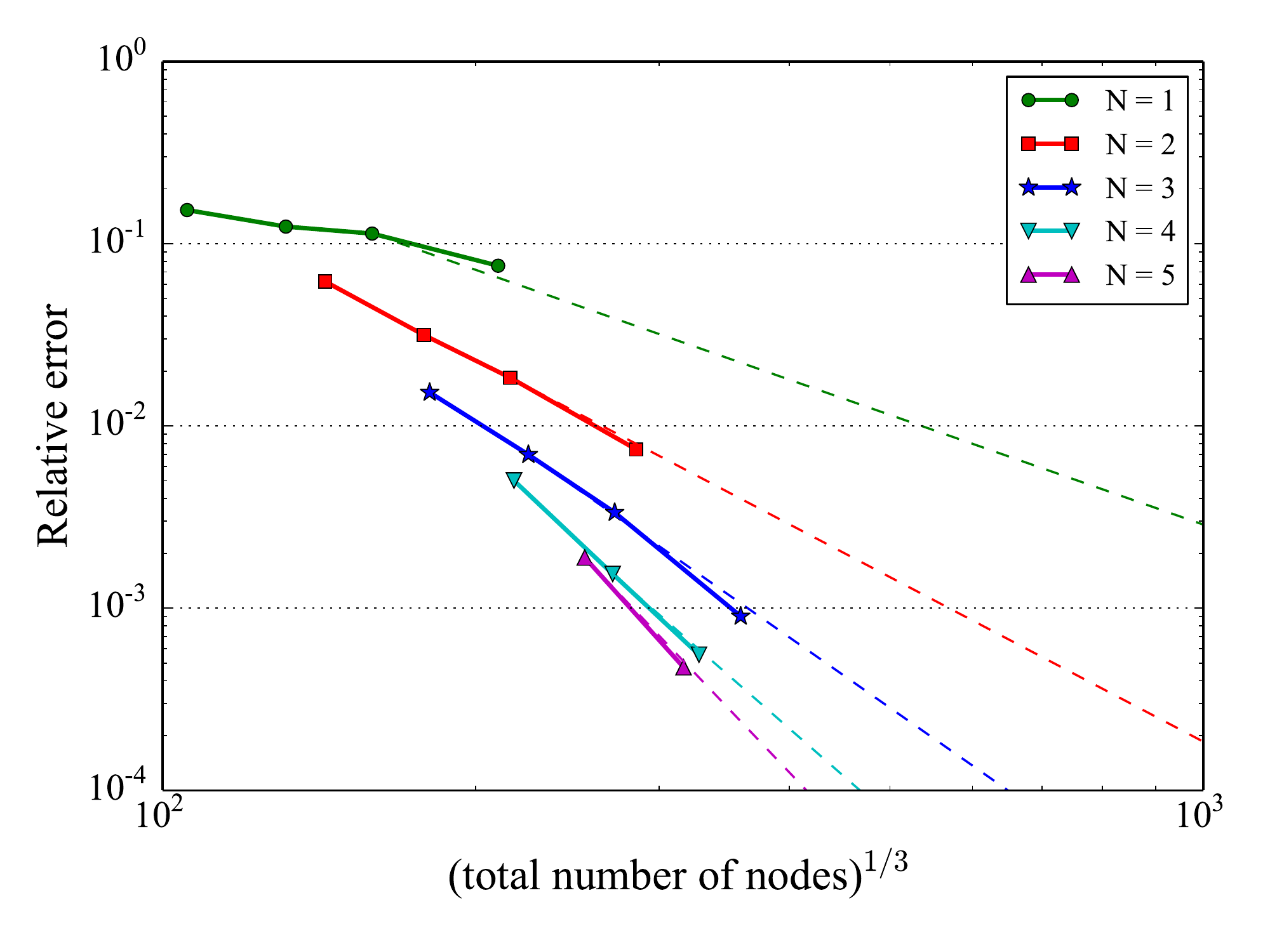} \vspace{-0.3cm}
  \caption{
Numerical convergence of the numerical scheme for different polynomial degrees $N$ with the validation case.
For each degree, the relative error is plotted as a function of the cube root of the total number of node $N_pK$.
Dashed lines correspond to $h^{p+1}$ convergence.
  }
  \label{fig:case1:numConv}
\end{figure}

The convergence of the error for polynomial basis functions with different degrees is shown in figure \ref{fig:case1:numConv}.
For each degree $N$, relative errors are obtained with several meshes (with $294\:508$, $567\:071$, $1\:005\:575$ and $2\:320\:289$ tetrahedra).
The time-stepping scheme is used with only one time level and the time-step
\begin{eqnarray*}
  \Delta t = 0.15\:\min_k\left\{\frac{\ell_k}{(N+1)^2 c_k}\right\},
\end{eqnarray*}
where $\ell_k$ is the smallest perpendicular distance between a face and its opposite vertex for tetrahedron $k$.
As expected, the error curve is close to the classical $h^{p+1}$ convergence of upwind fluxes for all polynomial degrees.

%%% INTEREST OF HIGH ORDER

\begin{figure}[!t]
  \centering
  \includegraphics[width=0.55\textwidth,natwidth=576px,natheight=432px]{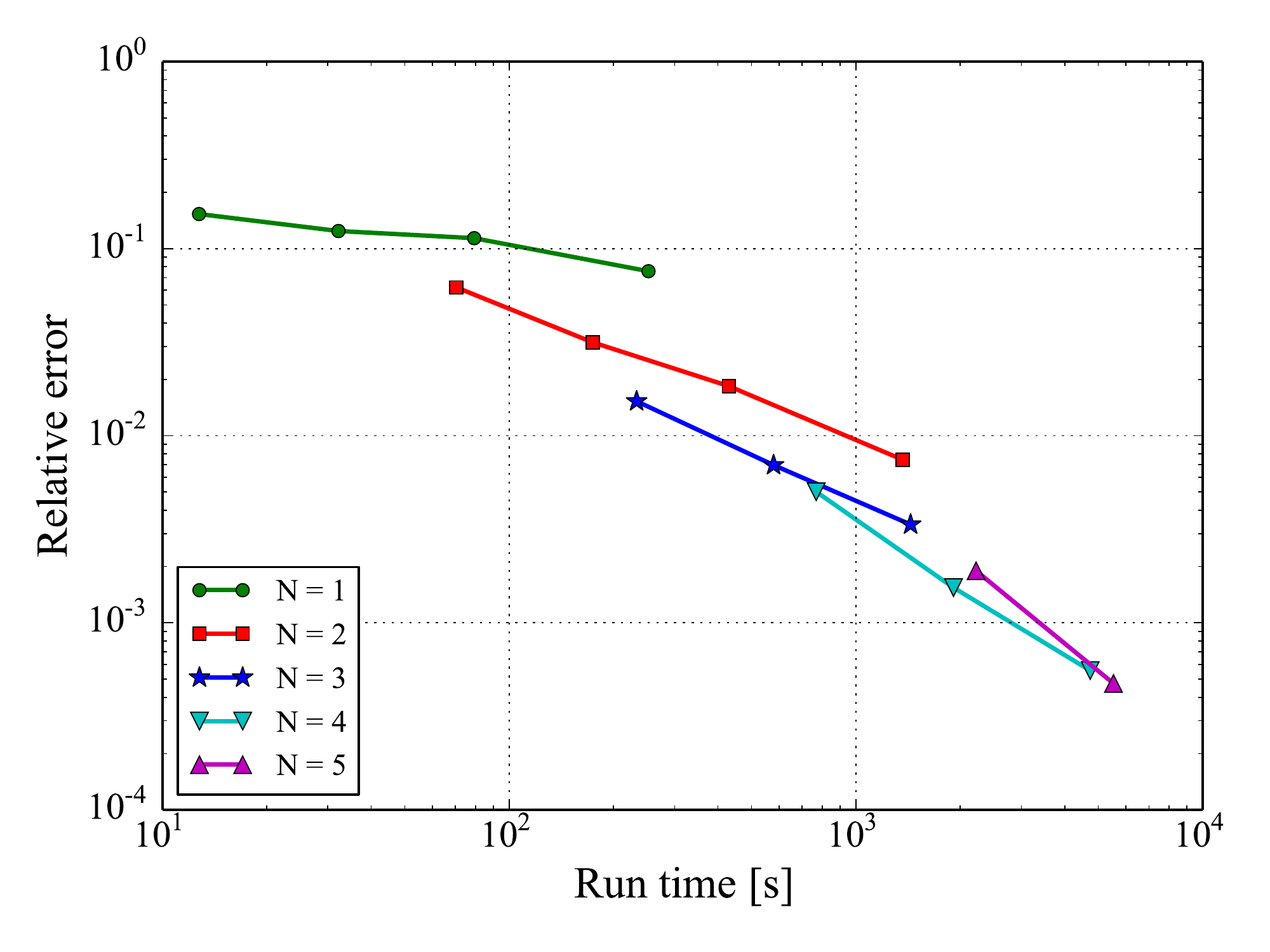} \vspace{-0.3cm}
  \caption{
Computational performance of the implementation for different polynomial degrees $N$ and 4 meshes with the validation case.
The run times are obtained with one single Nvidia K20m GPU and CUDA through OCCA.
For the highest degrees, simulations have not been performed with the finest meshes due to memory limitation of the GPU.}
  \label{fig:case1:compPerf}
\end{figure}

Figure \ref{fig:case1:compPerf} shows performance curves with a single Nvidia K20m GPU and the CUDA programming framework through OCCA.
It confirms the interest of hp-refinement strategy.
Indeed, an optimum polynomial degree exists depending on the desired accuracy.
For instance, to reach relative error $10^{-2}$, the third degree is more efficient than the second degree with a finer mesh.
By contrast, for the relative error $10^{-3}$, the fourth degree is worthwhile compared to the fifth with a coarser mesh.
Therefore, the best strategy to improve accuracy of solution is to refine the mesh and to increase the polynomial degree together.

%%% COMPARISON IMPLEMENTATION HARDWARE/METHOD

\begin{figure}[!t]
  \centering
  \includegraphics[width=0.55\textwidth,natwidth=576px,natheight=432px]{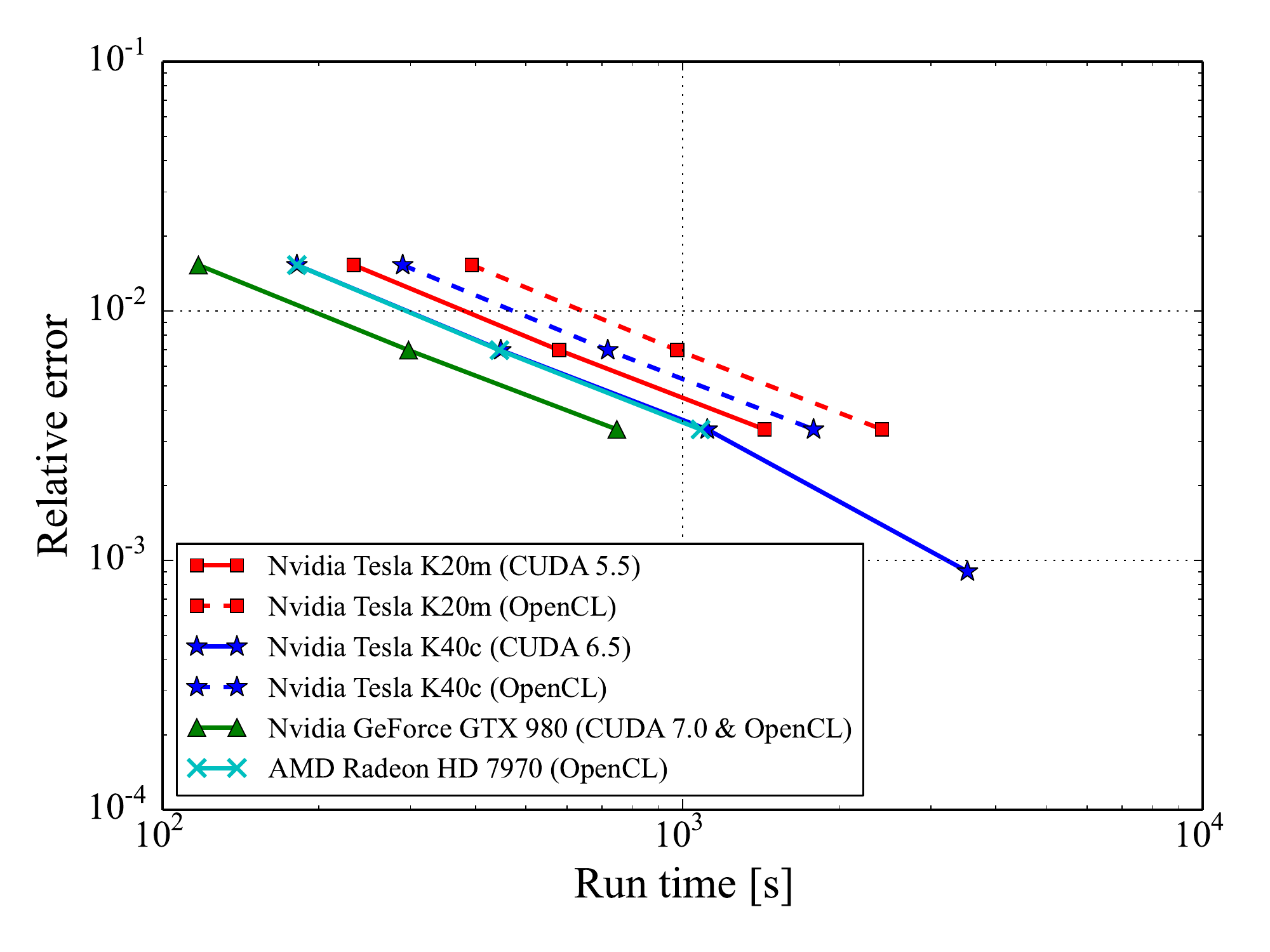} \vspace{-0.3cm}
  \caption{Comparison of run times obtained with different hardware devices. Third-degree polynomial basis functions are used with a single-rate time-stepping scheme.}
  \label{fig:case1:comparison}
\end{figure}

Finally, figure \ref{fig:case1:comparison} shows performance curves obtained with several GPUs: Nvidia Tesla K20m, Nvidia Tesla K40m, Nvidia GeForce GTX 980 and AMD Radeo HD 7970.
All the Nvidia GPUs have been tested with CUDA and Nvidia OpenCL.
The results are in accordance with the specifications of these GPUs.
The floating-point performances of both Tesla K40c and Radeon HD 7970 are similar.
Tesla K20m and GTX 980 have respectively the smallest and largest floating-point performances according to the specifications of the constructor.
For both Nvidia Tesla GPUs, the implementation is faster with CUDA than with OpenCL (speedup: $\sim 1.65$), while identical performances have been obtained with the GTX 980.
Only the Tesla K40c has a enough large memory to deal the finest mesh, for which 6.16GB must be stored on the GPU.

%% file: benchmpi.tex
\subsection{RTM Case for a Cluster of GPUs}
\label{sec:benchRTM}

This second benchmark deals with the complete RTM procedure for a larger geometry.
The classical imaging condition \eqref{eqn:imageP} and the new one based on characteristics \eqref{eqn:imageQ} are compared numerically.
Both scalability and memory transfers of our implementation are studied using up to 32 Nvidia K20 GPUs of TACC's Stampede cluster.

%%% DESCRIPTION

\begin{figure}
  \centering
  (a) Velocity model \\
  \includegraphics[width=0.65\textwidth,natwidth=1247px,natheight=820px]{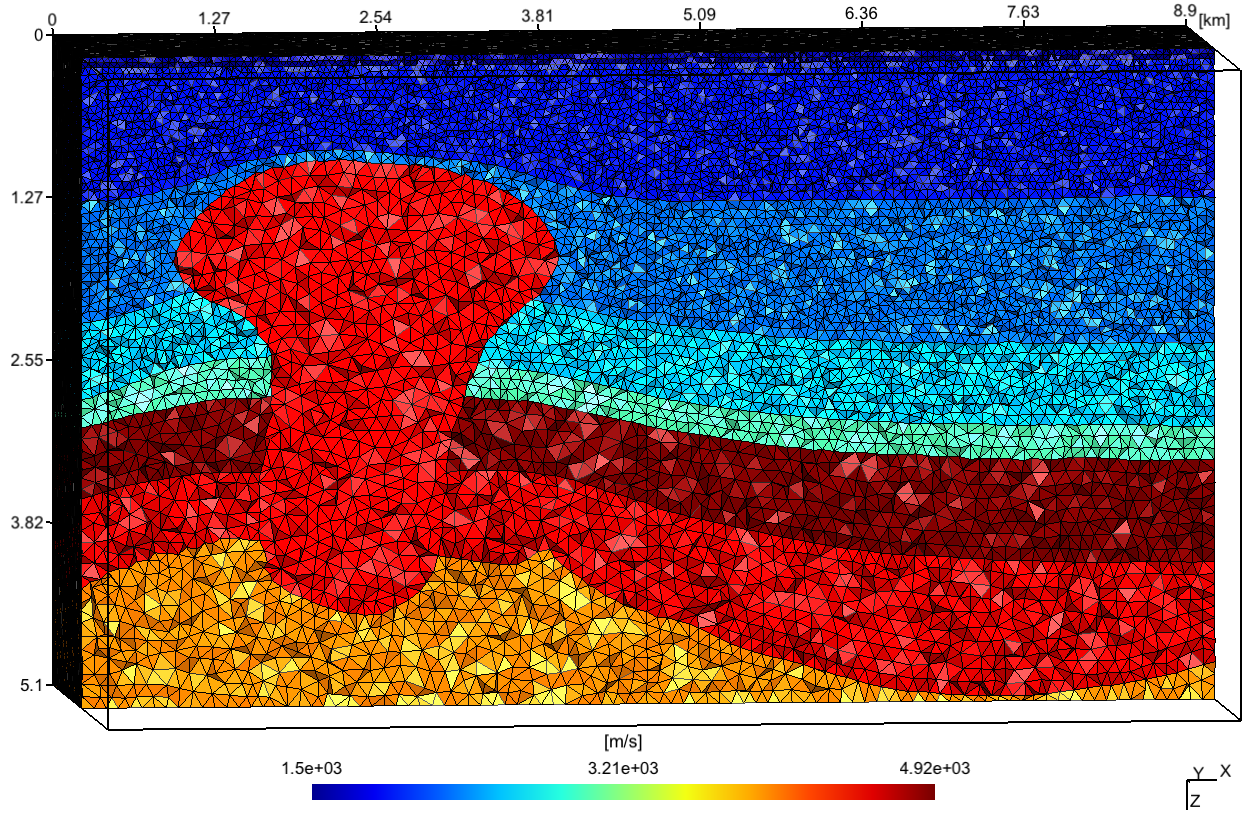} \vspace{0.2cm} \\
  (b) Time levels \\
  \includegraphics[width=0.65\textwidth,natwidth=1247px,natheight=820px]{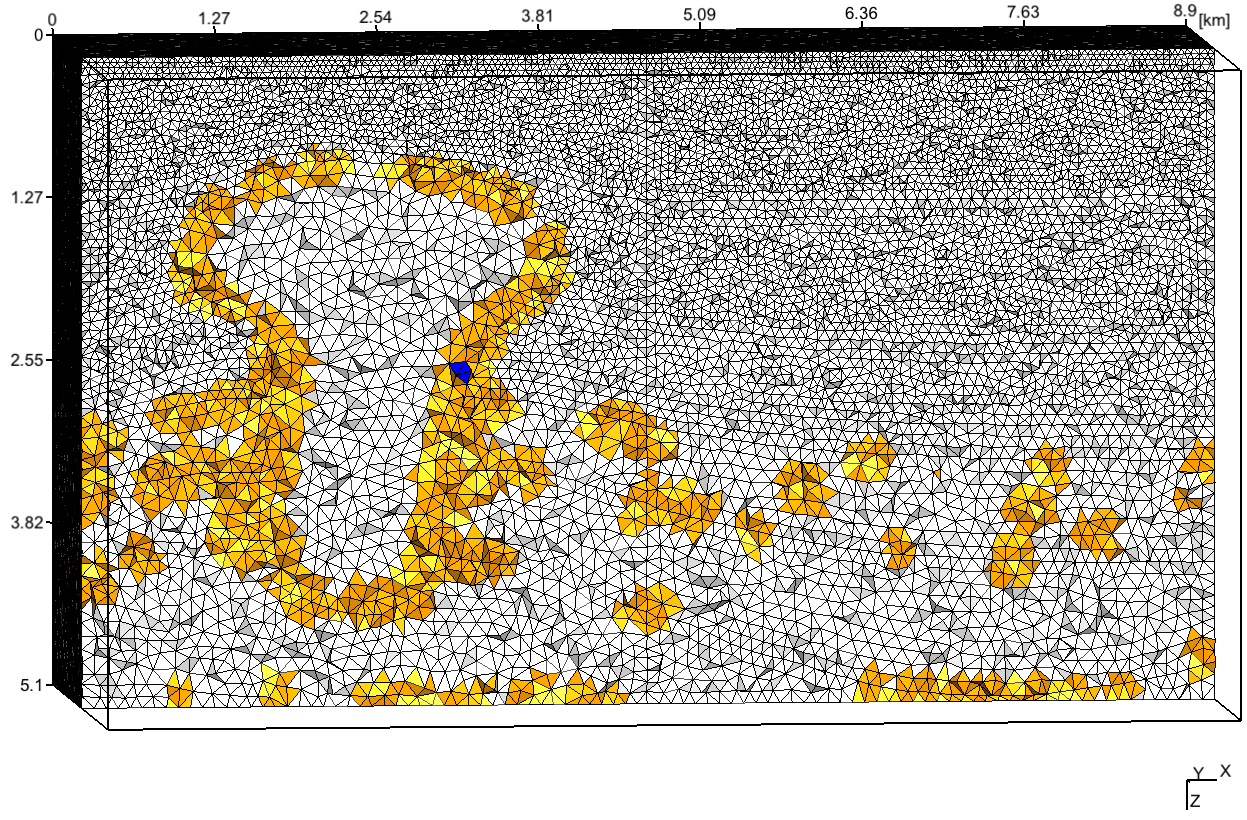}
  \caption{
Velocity model (a) and time levels (b) for the RTM benchmark.
Elements of the coarse, intermediate and fine levels are respectively white, orange and blue on figure (b).
The box indicates the total size of the computational domain.}
  \label{fig:RTM:description}
\end{figure}

The geometry consists in a salt dome embedded in a multi-layered domain of size $8.9\times4.44\times5.1\:\km^3$.
The top layer corresponds to sea water, while the other represent sediments and rocks.
The velocity model is shown in figure \ref{fig:RTM:description}(a).
The discretization consists of a tetrahedral mesh made of $2~787~335$ elements with a third-degree polynomial basis, which corresponds to $222~986~800$ discrete unknowns per simulation.
The mesh was generated by \cite{Kononov2012} and used by \cite{Minisini2013}.

For this case, the multi-rate time-stepping scheme is used with 3 time levels.
The coarse, intermediate and fine levels contain respectively $94.2\%$, $5.7\%$ and less than $0.1\%$ of elements.
The multi-rate time-stepping is thus particularly interesting here.
As shown in figure \ref{fig:RTM:description}b, most of the intermediate and fine elements are close to geometrical features (domain boundary and interfaces between layers).
Since the smallest elements are isolated in the mesh, the lumping strategy in the partition procedure allows us to avoid MPI exchanges for the finest level without penalizing the load balancing between MPI processes.

%%% IMAGING CONDITIONS  (classical OR with characteristic fields)
\subsubsection*{Synthetic Survey and Imaging Conditions}

A synthetic survey has been performed with 29 receiver gathers situated in the top layer of the domain.
The position of the first source is  $\vec{x}_s=(8\:\km,2.2\:\km,8\:\m)$, and the corresponding receivers are placed at depth $z_r=10\:\m$ on the grid $(x_r,y_r)\in[5\:\km,7.9\:\km]\times[2.1\:\km,2.3\:\km]$ with regular steps $25\:\m$ and $50\:\m$.
The same setting is used for the other gathers, but with positions moved $250\:\m$, $500\:\m$, \dots, $6.75\:\km$ and $7\:\km$ in direction $x$ negative.
In all cases, a Ricker pulse is emitted at the source with the frequency $50\:\Hz$, and the peak is generated at $t_p=0.1127\:\s$.
The total simulation time is $t_f=3.6127\:\s$.
The imaging conditions are computed over the period $[t_i,t_f]$ with $t_i=0.75\:\s$.
The backward run was only performed for that period.
The transparent condition of equation \eqref{eqn:ABC} is used at the domain boundary. The boundary values are saved and replayed for the source re-simulation, simultaneously with the reverse-time receiver wavefield computation.

%shot: xs=1000(50)8900, ys=2200, zs=8 m
%recv: xs-xr=100(25)3000, yr=2100(50)2300, zr=10 m (relative to each shot, marine type)
%recording: time=0(0.004)3.5 sec
%wavelet: 12 Hz Ricker centred around time 0

\begin{figure}[!t]
  \centering
  (a) Image with classical condition \\
  \includegraphics[width=0.65\textwidth,natwidth=1221px,natheight=704px]{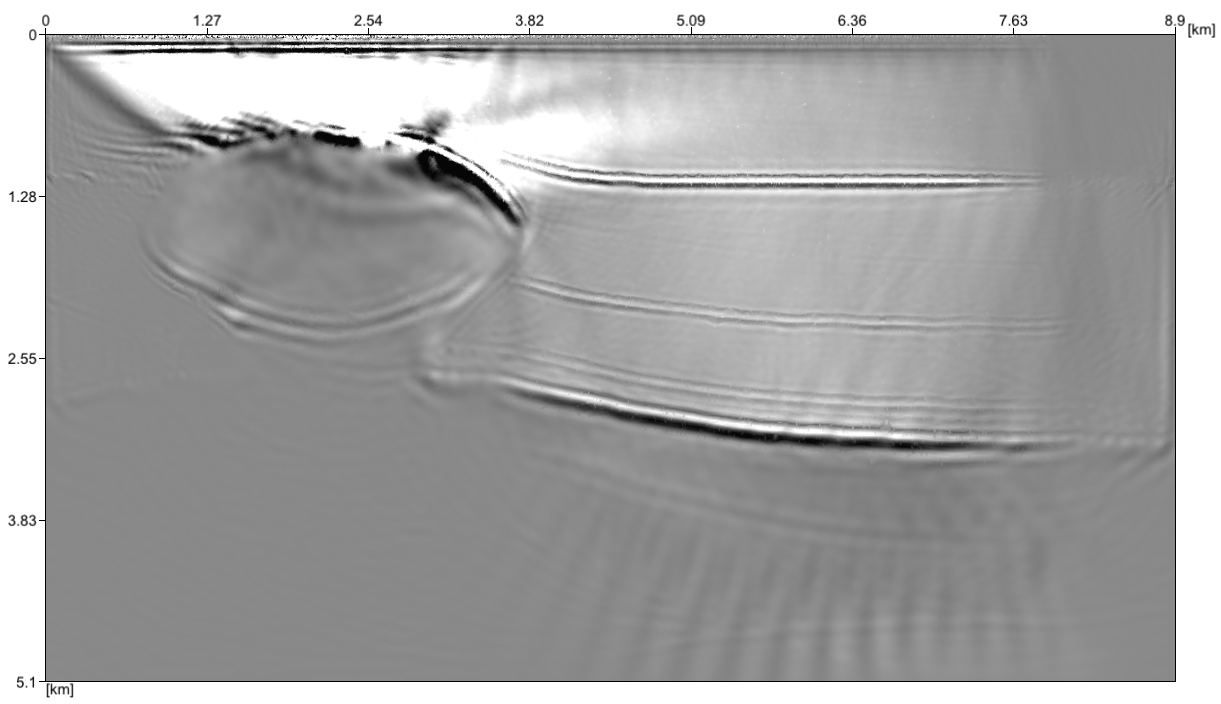} \\
  \vspace{0.3cm}
  (b) Image with characteristics-based condition \\
  \includegraphics[width=0.65\textwidth,natwidth=1221px,natheight=704px]{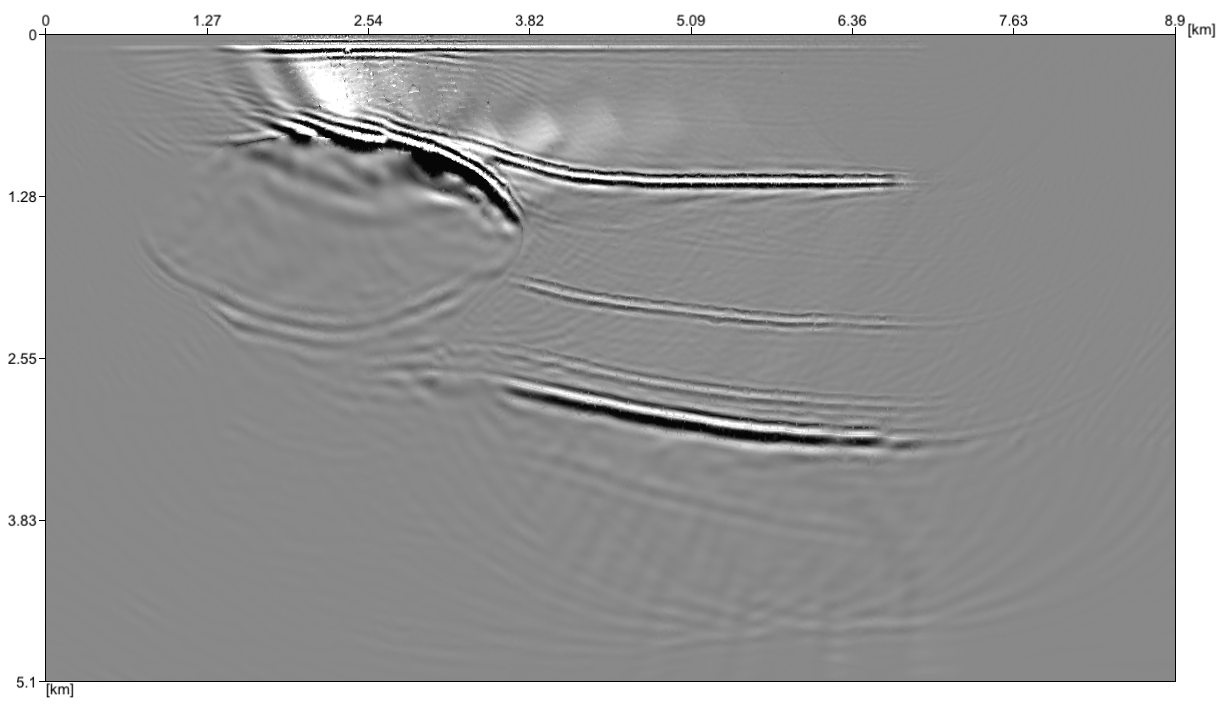}
  \caption{
Final images in the $x-z$ plane for the RTM benchmark obtained with the classical imaging condition (a) and the new one based on characteristics (b).
}
  \label{fig:RTM:images}
\end{figure}

\begin{figure}
  \centering
  (a) Pressure field $p$ \\
  \includegraphics[width=0.6\textwidth,natwidth=1221px,natheight=704px]{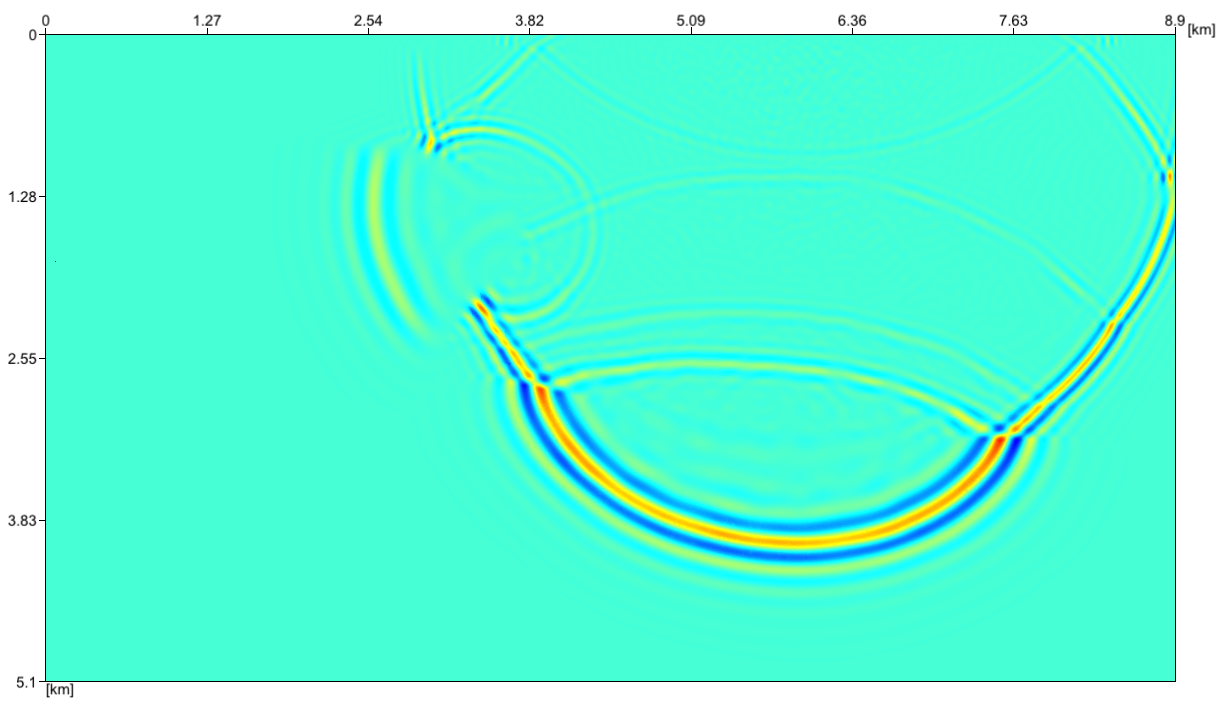} \\
  \vspace{0.3cm}
  (b) Upgoing characteristic field $q^+=p+c\rho\:\vec{e}_z\cdot\vec{v}$ \\
  \includegraphics[width=0.6\textwidth,natwidth=1221px,natheight=704px]{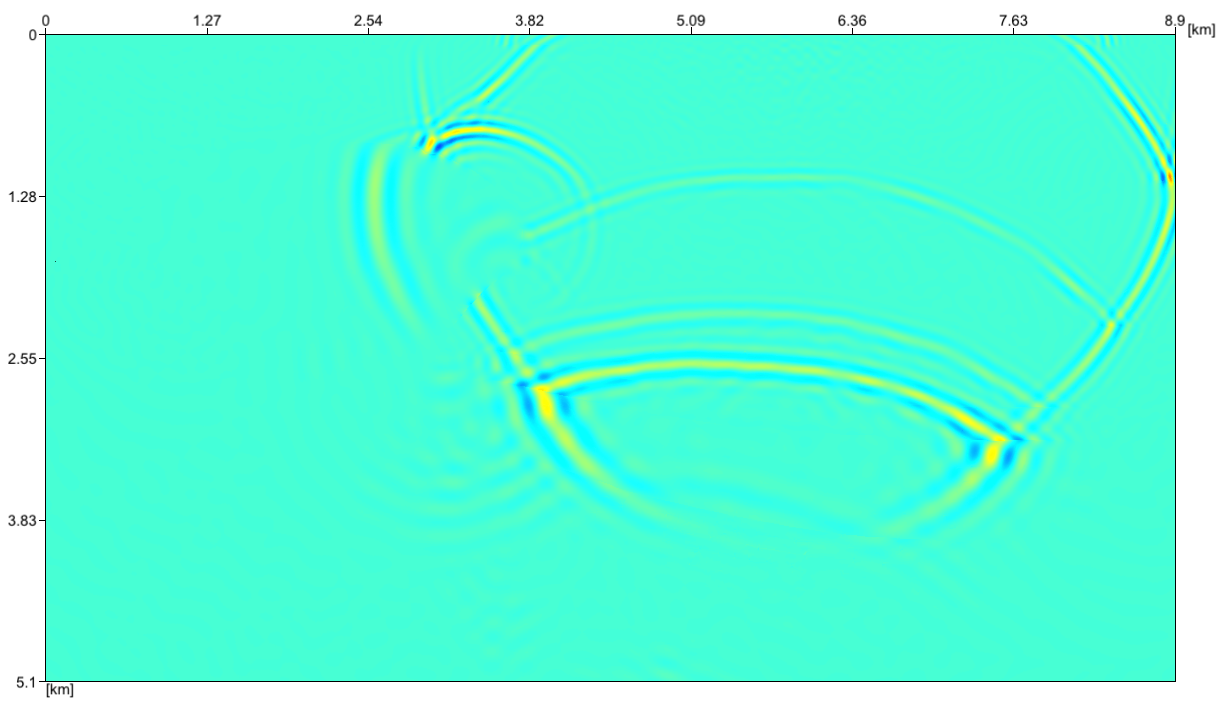} \\
  \vspace{0.3cm}
  (c) Downgoing characteristic field $q^-=p-c\rho\:\vec{e}_z\cdot\vec{v}$ \\
  \includegraphics[width=0.6\textwidth,natwidth=1221px,natheight=704px]{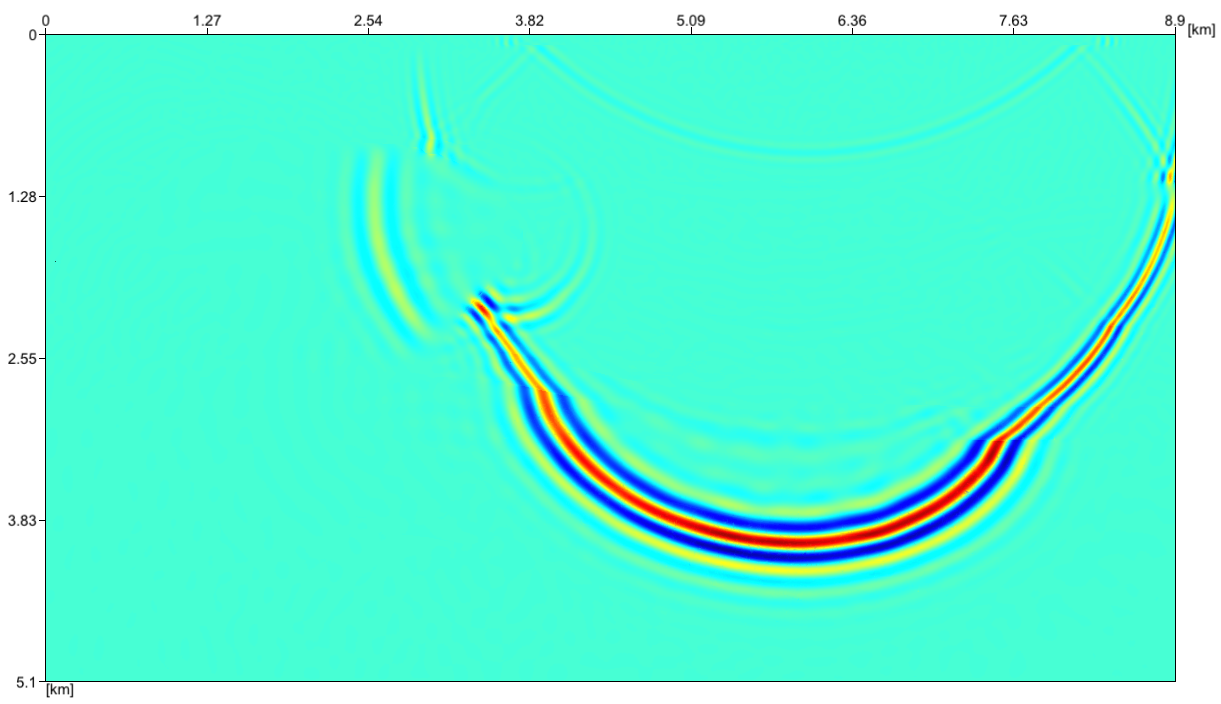}
  \caption{
Snapshots of pressure and upgoing/downgoing characteristics in the $x-z$ plane for one source simulation at one instant.
The same scale of color is used for all snapshots.
Wavefronts with a vertical propagation direction are decomposed into upgoing and downgoing components, only present in the corresponding characteristic fields.
Horizontal waves are present in both characteristic fields.
For wavefronts with an oblique propagation direction, the decomposition depends on the angle with the vertical direction.
}
  \label{fig:RTM:waveDecomp}
\end{figure}

Figure \ref{fig:RTM:images} shows the final images obtained with the classic imaging condition \eqref{eqn:imageP} and the characteristics-based one \eqref{eqn:imageQ}. 
The main geometrical features of the reflectors are imaged with both conditions, but low-frequency noise contaminates the image obtained with the classic condition.
This noise is suppressed by the new characteristics-based condition, providing a cleaner image.
A similar observation has been made by \cite{Liu2011} with their condition based on a spatial Fourier transform to decompose the wavefields into up- and downgoing components.
In contrast to that work, we use vertical characteristics to separate up- and downgoing waves at a negligible supplementary cost compared to the classic condition.
This decomposition is illustrated in figure \ref{fig:RTM:waveDecomp}.
We should emphasize that wavefields can be straightforwardly decomposed locally along any oblique direction by choosing characteristics with the corresponding direction vector instead of $\vec{e}_z$ in formula \eqref{eqn:charact}.

%%% COMPUTATIONAL PERFORMANCE
\subsubsection*{Memory and Computational Performance}

For each shot, the RTM procedure has been performed using 32 GPUs of TACC's Stampede cluster, and the gmsh software \citep{Geuzaine2009} was used for post-processing operations.
14.5GB of data must be allocated on GPUs, fairly well distributed between GPUs (from 353MB to 498MB by GPU).
Boundary data are here saved only on RAM (no RAM-HDD transfers).
They requires 240GB in total or, on average, 7.5GB by compute node.
However, the distribution between the compute nodes is not well balanced: one compute node does not store data (the corresponding sub-domain does not touch the domain boundary) and one node must store 14.34GB.
This is fortunately not a problem because compute node are configured with 32GB of host memory and, as shown later in this section, latency due to memory transfers between host (RAM) and device (GPU memory) is quite well hidden by computations.

The run time for the complete procedure with one shot is $12.08$ min, which includes a preparatory phase of $1.75$ min (mesh partition, generation of lists, etc.), and respectively $4.21$ min and $6.12$ min for forward and backward phases.
The preparatory phase can be done only once for the complete survey, since it is the same for all shots.
Let us mention that the backward phase has been made only for the period $[t_i,t_f]$.

%%% SCALABILITY
\subsubsection*{Scalability and Memory Transfers}

\begin{figure}
  \centering
  (a) Scalability for forward phase \\
  \includegraphics[width=0.7\textwidth,natwidth=576px,natheight=432px]{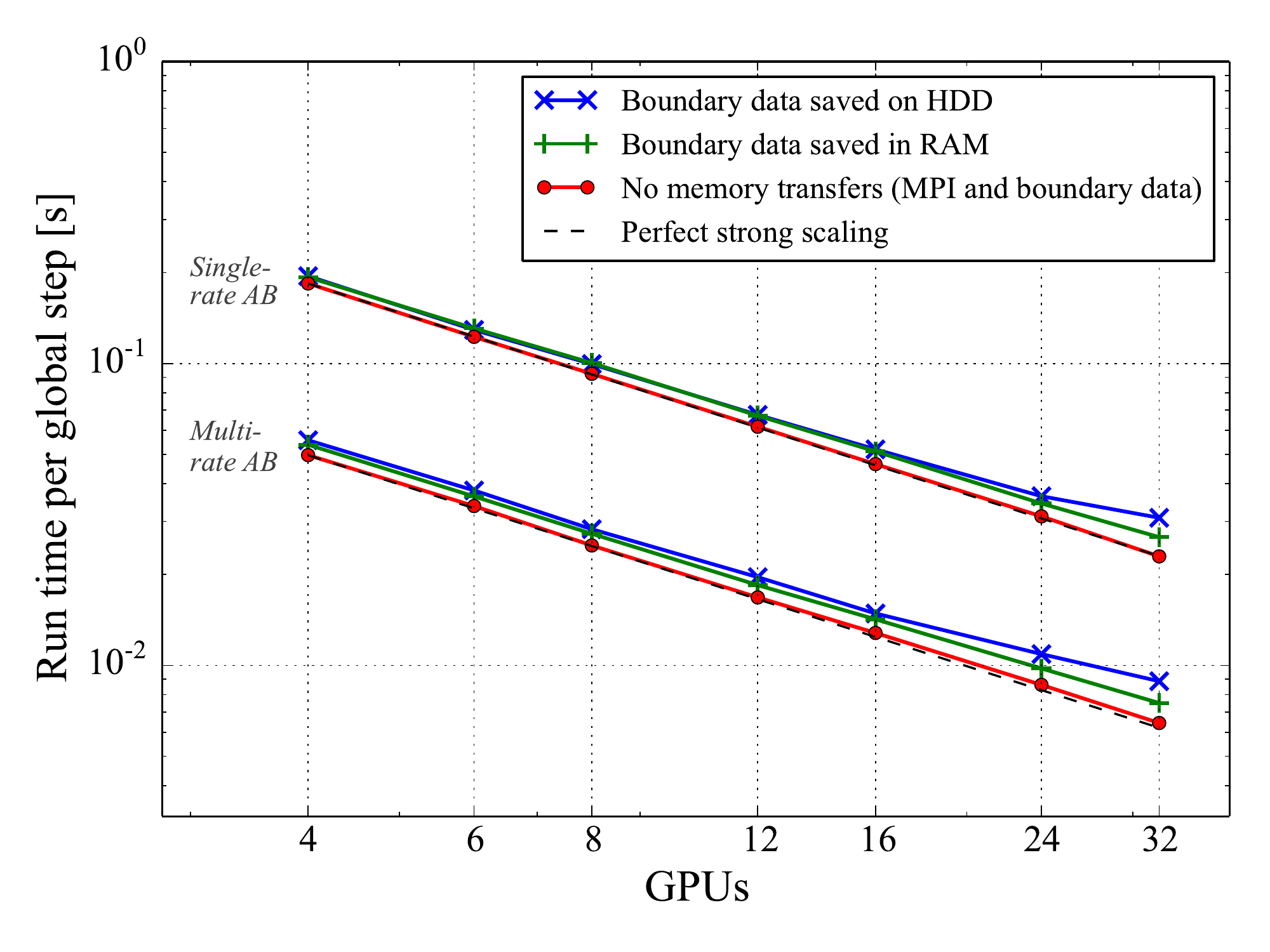} \\
  \bigskip
  (b) Scalability for backward phase \\
  \includegraphics[width=0.7\textwidth,natwidth=576px,natheight=432px]{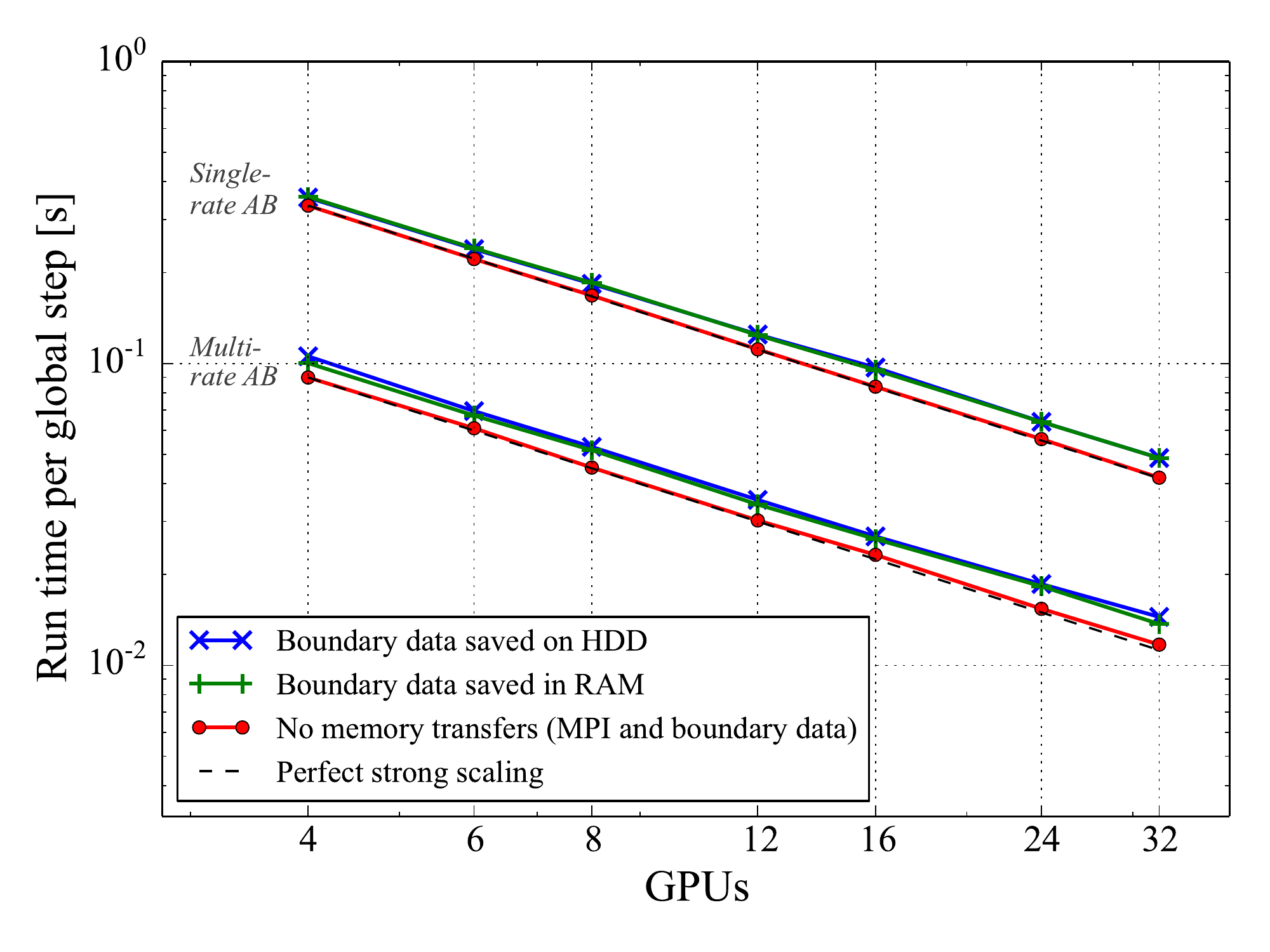}
  \caption{
Strong scalability for the RTM benchmark with 4 to 32 Nvidia K20 GPUs.
Run times are given separately for the forward phase (a) and the backward phase (b), both using the classical single-rate Adams-Bashforth (AB) scheme and using the multi-rate version with 3 time levels.
In order to study the influence of memory transfers on performance, we consider cases where the boundary data are stored on a HDD, where they are only stored in the RAM associated to each compute node, and where both MPI node-to-node communications and boundary data transfers are disabled.
  }
  \label{fig:RTM:scalability}
\end{figure}

Our implementation exhibits good strong scalability, despite high latency data transfers, as shown in figure \ref{fig:RTM:scalability}.
Run times are obtained using 4 to 32 GPUs, considering both single-rate and multi-rate time-stepping, and three alternative strategies to study memory transfers: boundary data are saved on a HDD, they are saved on the RAM associated to each compute node, and all slow communications are cut off (boundary data transfers and MPI communications).

When using the HDD storage for boundary data, the scalability is lost for the forward phase with 24 and 32 GPUs.
Let us recall that this storage requires supplementary RAM-HDD transfers, which are slower than transfers between memory device and RAM.
In this case, the RAM-HDD transfers, take too long to be hidden by the run time of kernels.
Fortunately, for a large enough number of GPUs, boundary data can be stored only in RAM, and  scalability is recovered.

Using multi-rate time-stepping instead of the single-rate version provides a speed up between $3.4$ and $3.6$, and preserves the scalability.
MPI transfers and device-host boundary data transfers are not perfectly hidden by computations because of blocking operations, such as the loading of traces received from MPI communications (see figure \ref{fig:flowchart}).
This however represent only few percent of the total run time, and does not impact the strong scalability.

%% file: conclusion.tex
We have presented a high-performance computational strategy for the RTM procedure with accelerator-aided clusters.
It combines efficient numerical schemes, a flexible programming approach, and a low memory storage/transfer strategy for RTM.
Our final implementation exhibits excellent strong scalability with massively parallel GPUs.
In addition, we have introduced a novel imaging condition that reduces noise in the final image, without significant extra computational cost in comparison with classical strategies.
We summarize hereafter key features contributing to performance:
\begin{itemize}

\item The spatial scheme, based on a nodal discontinuous Galerkin method, enables high-order convergence rates.
Since the weak element-to-element coupling and the dense algebraic operations required over each element, it is a suitable scheme for parallel computation on acceleration devices, especially with high-degree basis functions.

\item The levels-based multi-rate time-stepping scheme significantly improves run times for multi-scale cases with unstructured meshes, which are common in seismic imaging.
For parallel computations, mesh partition with METIS preserves good load balancing between cores thanks to strategies based on weights and lumping of elements.

\item The programming approach with MPI and OCCA provides a portable implementation that can be run on state-of-the-art computing architectures, and can be optimized for each of them.
While the results presented herein are obtained with kernels optimized for GPUs, tuning for alternative architectures requires to modify only a few parameters and, eventually, only a few code lines.

\item While RTM procedure generally requires massive data storage with slow I/O, the thin halo regions inherent in DG discretizations overcome this bottleneck.
Low storage requirements for DG boundary data allows halo trace data to be stored in RAM rather than relying on HDD, even for larger test cases.
In the worst cases, where available RAM is not large enough,  data exchanges between RAM and HDD can be performed asynchronously with a limited, possibly insignificant, increase of run time.

\item Several optimization strategies for data storage and data movement improve run time: locality of storage, specific granularity of storage to enable cache reuse, and hidden/asynchronous memory transfers for device-host and MPI node-to-node communications.

\end{itemize}
Future work includes the implementation of domain truncation methods to accurately deal with the artificial boundaries of the computational domain.
Techniques such as perfectly matched layers and complete radiation boundary conditions improve the solution close to artificial boundaries, but it would be challenging to incorporate them in our computational strategy without penalizing too much both run time and scalability.
We are also preparing a new implementation for hybrid meshes made of hexahedra, tetrahedra, prisms and pyramids.
DG implementations are particularly efficient for hexahedral elements, but mesh generators can only generate meshes dominated by hexahedra with a small amount of other elements.
In our RTM procedure, several compression strategies can be tested to reduce the size of boundary data.
Finally, in the future, we will propose cases involving more complicated physics, with anisotropic media and elastic waves, and generalization of the new characteristics-based imaging condition.

%% file: schemedgappendix.tex
The semi-discrete scheme is obtained from the formulation \eqref{eqn:weakFormP}--\eqref{eqn:weakFormV} by replacing the fields by their approximations \eqref{eqn:appFieldP}--\eqref{eqn:appFieldV}, and by using the Lagrange polynomial functions as test functions. For each node $n$ of element $k$, we then have
\begin{eqnarray*}
  \sum_{n=1}^{N_p} M_{k,mn} \ddiff{\disc{p}_{k,n}}{t}
    &=& - \sum_{i=1}^{3} \sum_{n=1}^{N_p} S_{k,i,mn}\:\rho_k c_k^2\:\disc{v}_{i,k,n}
       - \sum_{f=1}^{N_f} \sum_{n'=1}^{N_{fp}} M_{k,f,mn'}\:P_{k,n'}^p, \\
  \sum_{n=1}^{N_p} M_{k,mn} \ddiff{\disc{v}_{j,k,n}}{t}
    &=& - \sum_{n=1}^{N_p} S_{k,j,mn}\:\frac{1}{\rho_k}\:\disc{p}_{k,n}
       - \sum_{f=1}^{N_f} \sum_{n'=1}^{N_{fp}} M_{k,f,mn'}\:P_{k,n'}^{v_j},
\end{eqnarray*}
with $j=1,2,3$. In this system, $N_f$ is the number of faces by element, $N_{fp}$ is the number of nodes by face, $n'$ is the face node index, and the corresponding node index in the list of all element nodes is denoted with $n(n')$. We have defined the penalties evaluated at each face node as
\begin{eqnarray*}
  P_{k,f,n'}^p
    &=& \rho_k c_k^2
       \left(\left(\sum_{i=1}^3 n_{i,k,f}
           \left(\disc{v}_{i,k,n(n')}^\star-\disc{v}_{i,k,n(n')}\right)\right)
         - \tau_p \left(\disc{p}_{k,n(n')}^\star-\disc{p}_{k,n(n')}\right)\right), \\
  P_{k,f,n'}^{v_j}
    &=& \frac{n_{j,k,f}}{\rho_k}
       \left(\left(\disc{p}_{k,n(n')}^\star-\disc{p}_{k,n(n')}\right)
         - \tau_{\vec{v}}\:\sum_{i=1}^3 n_{i,k,f}
         \left(\disc{v}_{i,k,n(n')}^\star-\disc{v}_{i,k,n(n')}\right)\right),
\end{eqnarray*}
where $n_{i,k,f}$ is the $i^\text{th}$ Cartesian component of the outward unit normal of the face $\partial\D_{k,f}$. The star subscript denotes the nodal values on the side of the neighboring element. The components of the so-called local mass, local stiffness and local face mass matrices are respectively
\begin{eqnarray*}
  M_{k,mn} &=& \int_{\D_k}\ell_{k,m}\ell_{k,n}\:d\vec{x}, \\
  S_{k,i,mn} &=& \int_{\D_k}\ell_{k,m}\ddiff{\ell_{k,n}}{x_i}\:d\vec{x}, \\
  M_{k,f,mn'} &=& \int_{\partial\D_{k,f}}\ell_{k,m}\ell_{k,n(n')}\:d\vec{x}.
\end{eqnarray*}
Inverting the local mass matrix, the system can then be written as
\begin{eqnarray*}
  \ddiff{\vec{q}_k^p}{t}
    &=& - \rho_k c_k^2\:\sum_{i=1}^{3} \mat{M}_k^{-1} \mat{S}_{k,i}\:\vec{q}_k^{v_i}
    - \sum_{f=1}^{N_f} \mat{M}_k^{-1} \mat{M}_{k,f}\:\vec{P}_{k,f}^p, \\
  \ddiff{\vec{q}_k^{v_j}}{t}
    &=& - \frac{1}{\rho_k}\:\mat{M}_k^{-1} \mat{S}_{k,j}\:\vec{q}_k^p
    - \sum_{f=1}^{N_f} \mat{M}_k^{-1} \mat{M}_{k,f}\:\vec{P}_{k,f}^{v_j},
\end{eqnarray*}
where $\vec{q}_{k}^p$ and $\vec{q}_{k}^{v_j}$ contain the discrete unknowns associated to each field for all nodes of element $k$. The vectors $\vec{P}_{k,f}^p$ and $\vec{P}_{k,f}^{v_j}$ contain the penalties for all face nodes of face $f$. $\mat{M}_k$ and $\mat{S}_{k,i}$ are $N_p\times N_p$ matrices, while $\mat{M}_{k,f}$ is a $N_p\times N_{fp}$ matrix.

Since an affine transformation connects each element to a reference element, the local matrices can be expressed in terms of reference matrices that are the same for each element, combined with scaling and linear combinations. Denoting the transformation from the reference cell $\mathsf{I}$ to each cell $\D_k$ with $\boldsymbol{\Psi}_k : \vec{r}\in\mathsf{I} \rightarrow \vec{x}\in\D_k$, we have
\begin{eqnarray*}
  \vec{M}_{k} &=& J_k\:\vec{M}_{\mathsf{I}}, \\
  \vec{S}_{k,i} &=& J_k \sum_{j=1}^3 \diff{\Psi_{k,i}}{r_j} \vec{S}_{\mathsf{I},j},
\end{eqnarray*}
where $J_k = \det(\partial\boldsymbol{\Psi}_k/\partial\vec{r})$ is the Jacobian of the transformation. Similarly, each face is related to a reference face with an \textit{ad hoc} affine transformation. This permits to express all local face mass matrices in term of a reference two-dimensional mass matrix, involving the Jacobian of this transformation $J_{k,f}$ as well as a $N_p\times N_{fp}$ matrix which connects the face node indices with the node indices in the list of all nodes.

For each element, the system of equations can finally be written in the convenient form
\begin{eqnarray*}
  \ddiff{\vec{q}_k^p}{t}
    &=& - \rho_k c_k^2\:\sum_{i=1}^{3} \sum_{j=1}^3 \diff{\Psi_{k,i}}{r_j} \vec{D}_{j} \vec{q}_k^{v_i}
    - \sum_{f=1}^{N_f} \frac{J_{k,f}}{J_k} \mat{L}_{f}\vec{P}_{k,f}^p, \\
  \ddiff{\vec{q}_k^{v_i}}{t}
    &=& - \frac{1}{\rho_k} \sum_{j=1}^3 \diff{\Psi_{k,i}}{r_j} \vec{D}_{j}\vec{q}_k^p
    - \sum_{f=1}^{N_f} \frac{J_{k,f}}{J_k} \mat{L}_{f}\vec{P}_{k,f}^{v_i},
\end{eqnarray*}
where $\mat{D}_{j}$ are the differentiation matrices and $\mat{L}_{f}$ are the lifting matrices for the reference element. Defining the geometric factors
\begin{eqnarray*}
  g_{k,i,j}^\text{vol} &=& \diff{\Psi_{k,i}}{r_j}, \\
  g_{k,f}^\text{sur} &=& \frac{J_{k,f}}{J_k},
\end{eqnarray*}
one obtains equations \eqref{eqn:dgLocalSchemeP}-\eqref{eqn:dgLocalSchemeU}.

%% file: main.bbl
\begin{thebibliography}{55}
\providecommand{\natexlab}[1]{#1}
\providecommand{\url}[1]{\texttt{#1}}
\expandafter\ifx\csname urlstyle\endcsname\relax
  \providecommand{\doi}[1]{doi: #1}\else
  \providecommand{\doi}{doi: \begingroup \urlstyle{rm}\Url}\fi

\bibitem[Abdelkhalek et~al.(2012)Abdelkhalek, Calandra, Coulaud, Latu, and
  Roman]{Abdelkhalek2012}
R.~Abdelkhalek, H.~Calandra, O.~Coulaud, G.~Latu, and J.~Roman.
\newblock Fast seismic modeling and reverse time migration on a graphics
  processing unit cluster.
\newblock \emph{Concurrency and Computation: Practice and Experience},
  24\penalty0 (7):\penalty0 739--750, 2012.

\bibitem[Baldassari et~al.(2011)Baldassari, Barucq, Calandra, and
  Diaz]{Baldassari2011}
C.~Baldassari, H.~Barucq, H.~Calandra, and J.~Diaz.
\newblock Numerical performances of a hybrid local-time stepping strategy
  applied to the reverse time migration.
\newblock \emph{Geophysical Prospecting}, 59\penalty0 (5):\penalty0 907--919,
  2011.

\bibitem[Baldassari et~al.(2012)Baldassari, Barucq, Calandra, Denel, and
  Diaz]{Baldassari2012}
C.~Baldassari, H.~Barucq, H.~Calandra, B.~Denel, and J.~Diaz.
\newblock Performance analysis of a high-order discontinuous {G}alerkin method,
  application to the reverse time migration.
\newblock \emph{Communications in Computational Physics}, 11\penalty0
  (2):\penalty0 660--673, 2012.

\bibitem[Baysal et~al.(1983)Baysal, Kosloff, and Sherwood]{Baysal1983}
E.~Baysal, D.~Kosloff, and J.~Sherwood.
\newblock {Reverse time migration}.
\newblock \emph{Geophysics}, 48\penalty0 (11):\penalty0 1514--1524, 1983.
\newblock \doi{10.1190/1.1441434}.

\bibitem[Bleistein et~al.(2001)Bleistein, Cohen, and Stockwell]{Bleistein2001}
N.~Bleistein, J.~K. Cohen, and J.~W. Stockwell.
\newblock \emph{Mathematics of multidimensional seismic imaging, migration, and
  inversion}, volume~13.
\newblock Springer, 2001.

\bibitem[Chin-Joe-Kong et~al.(1999)Chin-Joe-Kong, Mulder, and
  Van~Veldhuizen]{Chin1999}
M.~J.~S. Chin-Joe-Kong, W.~A. Mulder, and M.~Van~Veldhuizen.
\newblock Higher-order triangular and tetrahedral finite elements with mass
  lumping for solving the wave equation.
\newblock \emph{Journal of Engineering Mathematics}, 35\penalty0 (4):\penalty0
  405--426, 1999.

\bibitem[Claerbout(1971)]{Claerbout1971}
J.~F. Claerbout.
\newblock Toward a unified theory of reflector mapping.
\newblock \emph{Geophysics}, 36\penalty0 (3):\penalty0 467--481, 1971.

\bibitem[Claerbout(1985)]{Claerbout1985}
J.~F. Claerbout.
\newblock \emph{Imaging the {E}arth's interior}.
\newblock Blackwell Scientific Publications, Inc., 1985.

\bibitem[Clapp(2009)]{Clapp2009}
R.~G. Clapp.
\newblock Reverse time migration with random boundaries.
\newblock In \emph{Proceedings of the 79th SEG Annual Meeting}, volume~28,
  pages 2809--2813, 2009.

\bibitem[Cohen et~al.(2001)Cohen, Joly, Roberts, and Tordjman]{Cohen2001}
G.~Cohen, P.~Joly, J.~E. Roberts, and N.~Tordjman.
\newblock Higher order triangular finite elements with mass lumping for the
  wave equation.
\newblock \emph{SIAM Journal on Numerical Analysis}, 38\penalty0 (6):\penalty0
  2047--2078, 2001.

\bibitem[Collis et~al.(2010)Collis, Ober, and van Bloemen~Waanders]{Collis2010}
S.~S. Collis, C.~C. Ober, and B.~G. van Bloemen~Waanders.
\newblock Unstructured discontinuous galerkin for seismic inversion.
\newblock In \emph{Proceedings of the 80th SEG Annual Meeting}, 2010.

\bibitem[Dumbser and K{\"a}ser(2006)]{Dumbser2006}
M.~Dumbser and M.~K{\"a}ser.
\newblock An arbitrary high-order discontinuous {G}alerkin method for elastic
  waves on unstructured meshes {--} {II}. {T}he three-dimensional isotropic
  case.
\newblock \emph{Geophysical Journal International}, 167\penalty0 (1):\penalty0
  319--336, 2006.

\bibitem[Dumbser and K{\"a}ser(2009)]{Dumbser2009}
M.~Dumbser and M.~K{\"a}ser.
\newblock A p-adaptive discontinuous {G}alerkin method with local time steps
  for computational seismology.
\newblock In \emph{High Performance Computing in Science and Engineering,
  Garching/Munich 2007}, pages 569--584. Springer, 2009.

\bibitem[Dussaud et~al.(2008)Dussaud, Symes, Williamson, Lemaistre, Singer,
  Denel, and Cherrett]{Dussaud2008}
E.~Dussaud, W.~W. Symes, P.~Williamson, L.~Lemaistre, P.~Singer, B.~Denel, and
  A.~Cherrett.
\newblock Computational strategies for reverse-time migration.
\newblock In \emph{Proceedings of the 78th SEG Annual Meeting}, page 2267,
  November 2008.

\bibitem[Etgen and Michelena(2010)]{Etgen2010}
J.~Etgen and R.~Michelena.
\newblock Introduction to this special section: Reverse time migration.
\newblock \emph{The Leading Edge}, 29\penalty0 (11):\penalty0 1363--1363, 2010.
\newblock \doi{10.1190/1.3517307}.

\bibitem[Etienne et~al.(2010)Etienne, Chaljub, Virieux, and
  Glinsky]{Etienne2010}
V.~Etienne, E.~Chaljub, J.~Virieux, and N.~Glinsky.
\newblock An hp-adaptive discontinuous {G}alerkin finite-element method for
  {3-D} elastic wave modelling.
\newblock \emph{Geophysical Journal International}, 183\penalty0 (2):\penalty0
  941--962, 2010.

\bibitem[Fuhry et~al.(2014)Fuhry, Giuliani, and Krivodonova]{Fuhry2014}
M.~Fuhry, A.~Giuliani, and L.~Krivodonova.
\newblock Discontinuous {G}alerkin methods on graphics processing units for
  nonlinear hyperbolic conservation laws.
\newblock \emph{International Journal for Numerical Methods in Fluids},
  76\penalty0 (12):\penalty0 982--1003, 2014.

\bibitem[Gandham et~al.(2015)Gandham, Medina, and Warburton]{Gandham2015}
R.~Gandham, D.~Medina, and T.~Warburton.
\newblock {GPU} accelerated discontinuous {G}alerkin methods for shallow water
  equations.
\newblock \emph{Communications in Computational Physics}, 2015.
\newblock to appear.

\bibitem[Geuzaine and Remacle(2009)]{Geuzaine2009}
C.~Geuzaine and J.-F. Remacle.
\newblock Gmsh: A {3-D} finite element mesh generator with built-in pre-and
  post-processing facilities.
\newblock \emph{International Journal for Numerical Methods in Engineering},
  79\penalty0 (11):\penalty0 1309--1331, 2009.

\bibitem[Godel et~al.(2010)Godel, Schomann, Warburton, and Clemens]{Godel2010}
N.~Godel, S.~Schomann, T.~Warburton, and M.~Clemens.
\newblock {GPU} accelerated {A}dams-{B}ashforth multirate discontinuous
  {G}alerkin {FEM} simulation of high-frequency electromagnetic fields.
\newblock \emph{Magnetics, IEEE Transactions on}, 46\penalty0 (8):\penalty0
  2735--2738, 2010.

\bibitem[Hesthaven and Warburton(2002)]{Hesthaven2002}
J.~S. Hesthaven and T.~Warburton.
\newblock Nodal high-order methods on unstructured grids: {I}. {T}ime-domain
  solution of {M}axwell's equations.
\newblock \emph{Journal of Computational Physics}, 181\penalty0 (1):\penalty0
  186--221, 2002.

\bibitem[Hesthaven and Warburton(2007)]{Hesthaven2007}
J.~S. Hesthaven and T.~Warburton.
\newblock \emph{Nodal discontinuous {G}alerkin methods: algorithms, analysis,
  and applications}, volume~54.
\newblock Springer, 2007.

\bibitem[Hu and McMechan(1987)]{Hu1987}
L.-Z. Hu and G.~A. McMechan.
\newblock Wave-field transformations of vertical seismic profiles.
\newblock \emph{Geophysics}, 52\penalty0 (3):\penalty0 307--321, 1987.

\bibitem[Kl\"ockner et~al.(2009)Kl\"ockner, Warburton, Bridge, and
  Hesthaven]{Klockner2009}
A.~Kl\"ockner, T.~Warburton, J.~Bridge, and J.~S. Hesthaven.
\newblock Nodal discontinuous {G}alerkin methods on graphics processors.
\newblock \emph{Journal of Computational Physics}, 228\penalty0 (21):\penalty0
  7863--7882, 2009.

\bibitem[Kl\"ockner et~al.(2012)Kl\"ockner, Warburton, and
  Hesthaven]{Klockner2012}
A.~Kl\"ockner, T.~Warburton, and J.~S. Hesthaven.
\newblock Solving wave equations on unstructured geometries.
\newblock In W.~W. Hwu, editor, \emph{{GPU} Computing Gems Jade Edition},
  Applications of GPU Computing, pages 225--242. Morgan Kaufmann, 2012.

\bibitem[Kl\"ockner et~al.(2013)Kl\"ockner, Warburton, and
  Hesthaven]{Klockner2013}
A.~Kl\"ockner, T.~Warburton, and J.~S. Hesthaven.
\newblock High-order discontinuous {G}alerkin methods by {GPU} metaprogramming.
\newblock In D.~A. Yuen, L.~Wang, X.~Chi, L.~Johnsson, W.~Ge, and Y.~Shi,
  editors, \emph{GPU Solutions to Multi-scale Problems in Science and
  Engineering}, Lecture Notes in Earth System Sciences, pages 353--374.
  Springer Berlin Heidelberg, 2013.

\bibitem[Komatitsch and Tromp(1999)]{Komatitsch1999}
D.~Komatitsch and J.~Tromp.
\newblock Introduction to the spectral element method for three-dimensional
  seismic wave propagation.
\newblock \emph{Geophysical journal international}, 139\penalty0 (3):\penalty0
  806--822, 1999.

\bibitem[Komatitsch and Vilotte(1998)]{Komatitsch1998}
D.~Komatitsch and J.-P. Vilotte.
\newblock The spectral element method: an efficient tool to simulate the
  seismic response of {2D} and {3D} geological structures.
\newblock \emph{Bulletin of the seismological society of America}, 88\penalty0
  (2):\penalty0 368--392, 1998.

\bibitem[Komatitsch et~al.(2010)Komatitsch, Erlebacher, G\"oddeke, and
  Mich\'ea]{Komatitsch2010}
D.~Komatitsch, G.~Erlebacher, D.~G\"oddeke, and D.~Mich\'ea.
\newblock High-order finite-element seismic wave propagation modeling with
  {MPI} on a large {GPU} cluster.
\newblock \emph{Journal of Computational Physics}, 229\penalty0 (20):\penalty0
  7692--7714, 2010.

\bibitem[Kononov et~al.(2012)Kononov, Minisini, Zhebel, and
  Mulder]{Kononov2012}
A.~Kononov, S.~Minisini, E.~Zhebel, and W.~A. Mulder.
\newblock A {3D} tetrahedral mesh generator for seismic problems.
\newblock In \emph{Proceedings of the 74th EAGE Conference \& Exhibition}, page
  B006, June 2012.

\bibitem[Krebs et~al.(2014)Krebs, Collis, Downey, Ober, Overfelt, Smith, van
  Bloemen-Waanders, and Young]{Krebs2014}
J.~R. Krebs, S.~S. Collis, N.~J. Downey, C.~C. Ober, J.~R. Overfelt, T.~M.
  Smith, B.~G. van Bloemen-Waanders, and J.~G. Young.
\newblock Full wave inversion using a spectral-element discontinuous {G}alerkin
  method.
\newblock In \emph{Proceedings of the 76th EAGE Conference \& Exhibition},
  2014.

\bibitem[Lailly(1983)]{ref:lailly}
P.~Lailly.
\newblock The seismic inverse problem as a sequence of before stack migration.
\newblock In J.~B. Bednar, R.~Redner, E.~Robinson, and A.~Weglein, editors,
  \emph{Proceedings of the conference on inverse scattering: theory and
  applications}, pages 206--220, Philadelphia (PA), USA, 1983. SIAM.
\newblock ISBN 0-898-71190-8.

\bibitem[Liu et~al.(2011)Liu, Zhang, Morton, and Leveille]{Liu2011}
F.~Liu, G.~Zhang, S.~A. Morton, and J.~P. Leveille.
\newblock An effective imaging condition for reverse-time migration using
  wavefield decomposition.
\newblock \emph{Geophysics}, 76\penalty0 (1):\penalty0 S29--S39, 2011.

\bibitem[Liu et~al.(2013)Liu, Liu, Ren, and Meng]{Liu2013}
G.~Liu, Y.~Liu, L.~Ren, and X.~Meng.
\newblock {3D} seismic reverse time migration on {GPGPU}.
\newblock \emph{Computers \& Geosciences}, 59:\penalty0 17--23, 2013.

\bibitem[Liu et~al.(2012)Liu, Li, Liu, Tong, Liu, Wang, and Liu]{Liu2012}
H.~Liu, B.~Li, H.~Liu, X.~Tong, Q.~Liu, X.~Wang, and W.~Liu.
\newblock The issues of prestack reverse time migration and solutions with
  graphic processing unit implementation.
\newblock \emph{Geophysical Prospecting}, 60\penalty0 (5):\penalty0 906--918,
  2012.

\bibitem[{Loewenthal}(1983)]{Loewenthal1983}
D.~{Loewenthal}.
\newblock {Reversed time migration in spatial frequency domain}.
\newblock \emph{Geophysics}, 48:\penalty0 627, May 1983.
\newblock \doi{10.1190/1.1441493}.

\bibitem[McMechan(1983)]{Mcmechan1983}
G.~A. McMechan.
\newblock Migration by extrapolation of time-dependent boundary values.
\newblock \emph{Geophysical Prospecting}, 31\penalty0 (3):\penalty0 413--420,
  1983.

\bibitem[Medina et~al.(2014)Medina, St.{-}Cyr, and Warburton]{Medina2014}
D.~S. Medina, A.~St.{-}Cyr, and T.~Warburton.
\newblock {OCCA:} {A} unified approach to multi-threading languages.
\newblock 2014.

\bibitem[Mercerat and Glinsky(2015)]{Mercerat2015}
E.~D. Mercerat and N.~Glinsky.
\newblock A nodal high-order discontinuous {G}alerkin method for elastic wave
  propagation in arbitrary heterogeneous media.
\newblock \emph{Geophysical Journal International}, 201\penalty0 (2):\penalty0
  1101--1118, 2015.

\bibitem[Mich{\'e}a and Komatitsch(2010)]{Michea2010}
D.~Mich{\'e}a and D.~Komatitsch.
\newblock Accelerating a three-dimensional finite-difference wave propagation
  code using {GPU} graphics cards.
\newblock \emph{Geophysical Journal International}, 182\penalty0 (1):\penalty0
  389--402, 2010.

\bibitem[Minisini et~al.(2012)Minisini, Zhebel, Kononov, and
  Mulder]{Minisini2012}
S.~Minisini, E.~Zhebel, A.~Kononov, and W.~A. Mulder.
\newblock Efficiency comparison for continuous mass-lumped and discontinuous
  {G}alerkin finite-elements for {3D} wave propagation.
\newblock In \emph{Proceedings of the 74th EAGE Conference \& Exhibition}, page
  A004, June 2012.

\bibitem[Minisini et~al.(2013)Minisini, Zhebel, Kononov, and
  Mulder]{Minisini2013}
S.~Minisini, E.~Zhebel, A.~Kononov, and W.~A. Mulder.
\newblock Local time stepping with the discontinuous {G}alerkin method for wave
  propagation in {3D} heterogeneous media.
\newblock \emph{Geophysics}, 78\penalty0 (3):\penalty0 T67--T77, 2013.

\bibitem[Moczo et~al.(2011)Moczo, Kristek, Galis, Chaljub, and
  Etienne]{Moczo2011}
P.~Moczo, J.~Kristek, M.~Galis, E.~Chaljub, and V.~Etienne.
\newblock {3-D} finite-difference, finite-element, discontinuous-{G}alerkin and
  spectral-element schemes analysed for their accuracy with respect to {P}-wave
  to {S}-wave speed ratio.
\newblock \emph{Geophysical Journal International}, 187\penalty0 (3):\penalty0
  1645--1667, 2011.

\bibitem[Modave et~al.(2015)Modave, St-Cyr, Warburton, and Mulder]{Modave2015}
A.~Modave, A.~St-Cyr, T.~Warburton, and W.~A. Mulder.
\newblock Accelerated discontinuous {G}alerkin time-domain simulations for
  seismic wave propagation.
\newblock In \emph{Proceedings of the 77th EAGE Conference \& Exhibition},
  2015.

\bibitem[Mu et~al.(2013)Mu, Chen, and Wang]{Mu2013b}
D.~Mu, P.~Chen, and L.~Wang.
\newblock Accelerating the discontinuous {G}alerkin method for seismic wave
  propagation simulations using multiple {GPU}s with cuda and mpi.
\newblock \emph{Earthquake Science}, 26\penalty0 (6):\penalty0 377--393, 2013.

\bibitem[Nguyen and McMechan(2015)]{Nguyen2015}
B.~D. Nguyen and G.~A. McMechan.
\newblock Five ways to avoid storing source wavefield snapshots in {2D} elastic
  prestack reverse time migration.
\newblock \emph{Geophysics}, 80\penalty0 (1):\penalty0 S1--S18, 2015.

\bibitem[Symes(2007)]{Symes2007}
W.~W. Symes.
\newblock Reverse time migration with optimal checkpointing.
\newblock \emph{Geophysics}, 72\penalty0 (5):\penalty0 SM213--SM221, 2007.

\bibitem[Tarantola(1984)]{ref:tarantolaacou}
A.~Tarantola.
\newblock Inversion of seismic reflection data in the acoustic approximation.
\newblock \emph{Geophysics}, 49\penalty0 (8):\penalty0 1259--1266, 1984.
\newblock \doi{10.1190/1.1441754}.

\bibitem[Virieux et~al.(2011)Virieux, Calandra, and Plessix]{Virieux2011}
J.~Virieux, H.~Calandra, and R.-{\'E}. Plessix.
\newblock A review of the spectral, pseudo-spectral, finite-difference and
  finite-element modelling techniques for geophysical imaging.
\newblock \emph{Geophysical Prospecting}, 59\penalty0 (5):\penalty0 794--813,
  2011.

\bibitem[Warburton(2006)]{Warburton2006}
T.~Warburton.
\newblock An explicit construction of interpolation nodes on the simplex.
\newblock \emph{Journal of Engineering Mathematics}, 56\penalty0 (3):\penalty0
  247--262, 2006.

\bibitem[Warburton(2013)]{Warburton2013}
T.~Warburton.
\newblock A low-storage curvilinear discontinuous {G}alerkin method for wave
  problems.
\newblock \emph{SIAM Journal on Scientific Computing}, 35\penalty0
  (4):\penalty0 1987--2012, 2013.

\bibitem[Weiss and Shragge(2013)]{Weiss2013}
R.~M. Weiss and J.~Shragge.
\newblock Solving {3D} anisotropic elastic wave equations on parallel {GPU}
  devices.
\newblock \emph{Geophysics}, 78\penalty0 (2):\penalty0 F7--F15, 2013.

\bibitem[Whitmore(1983)]{Whitmore1983}
N.~D. Whitmore.
\newblock Iterative depth migration by backward time propagation.
\newblock In \emph{Proceedings of the 53th SEG Annual Meeting}, pages 382--385.
  Society of Exploration Geophysicists, 1983.

\bibitem[Yang et~al.(2014)Yang, Gao, and Wang]{Yang2014}
P.~Yang, J.~Gao, and B.~Wang.
\newblock {RTM} using effective boundary saving: A staggered grid {GPU}
  implementation.
\newblock \emph{Computers \& Geosciences}, 68:\penalty0 64--72, 2014.

\bibitem[Zhebel et~al.(2014)Zhebel, Minisini, Kononov, and Mulder]{Zhebel2014}
E.~Zhebel, S.~Minisini, A.~Kononov, and W.~A. Mulder.
\newblock A comparison of continuous mass-lumped finite elements with finite
  differences for {3-D} wave propagation.
\newblock \emph{Geophysical Prospecting}, 62\penalty0 (5):\penalty0 1111--1125,
  2014.

\end{thebibliography}
